# A Panoramic View of MXenes via a New Design Strategy

Noah Oyeniran[a], Oyshee Chowdhury[a], Chongze Hu[a*], Traian Dumitrica[b], Panchapakesan Ganesh[c], Jacek Jakowski[d], Zhongfang Chen[e], Raymond R. Unocic[f], Michael Naguib[g], Vincent Meunier[h], Yury Gogotsi[i], Paul R. C. Kent[d], Bobby G. Sumpter[c], Jingsong Huang[c*]

[a] *Department of Aerospace Engineering and Mechanics, The University of Alabama, Tuscaloosa, Alabama 35487, USA*

[b] *Department of Mechanical Engineering, University of Minnesota Twin Cities, Minneapolis, Minnesota 55455, USA*

[c] *Center of Nanophase Materials Sciences, Oak Ridge National Laboratory, Oak Ridge, Tennessee 37831, USA*

[d] *Computational Sciences and Engineering Division, Oak Ridge National Laboratory, Oak Ridge, Tennessee 37831, USA*

[e] *Department of Chemistry, University of Puerto Rico, Rio Piedras, San Juan, Puerto Rico 00931, USA*

[f] *Department of Materials Science and Engineering, North Carolina State University, Raleigh, North Caroline 27695, USA*

[g] *Department of Physics and Engineering Physics, Tulane University, New Orleans, Louisiana 70118, USA*

[h] *Department of Engineering Science and Mechanics, The Pennsylvania State University, University Park, Pennsylvania 16801, USA*

[i] *Department of Materials Science and Engineering, and A.J. Drexel Nanomaterials Institute, Drexel University, Philadelphia, Pennsylvania 19104, USA*

*Corresponding authors
Email address: hucz@ua.edu (Chongze Hu) and huangj3@ornl.gov (Jingsong Huang)

**Abstract**

Two-dimensional (2D) transition metal carbides and nitrides, known as MXenes, possess unique physical and chemical properties, enabling diverse applications in fields ranging from energy storage to communication, catalysis, sensing, healthcare, and beyond. The transition metal and nonmetallic atoms in MXenes can exhibit distinct coordination environments, potentially leading to a wide variety of 2D phases. Despite extensive research and significant advancements, a fundamental understanding of MXenes' phase diversity and its relationship with their hierarchical precursors, including intermediate MAX phases and parent bulk phases, remains limited. Using high-throughput modeling based on first-principles density functional theory, we unveil a wide range of MXenes and comprehensively evaluate their relative stabilities across a large chemical space. The key lies in considering both octahedral and trigonal prismatic coordination environments characteristic of various bulk phases. Through this comprehensive structural library of MXenes, we uncover a close alignment between the phase stability of MXenes and that of their hierarchical 3D counterparts. Building on this, we demonstrate a new design strategy where the atomic coordination environments in parent bulk phases can serve as reliable predictors for the design of MXenes, reducing reliance on intermediate MAX phases. Our study significantly expands the landscape of MXenes, at least doubling the number of possible structures.

**Introduction**

Layered transition metal carbides (TMC) or nitrides (TMN), also known as MXenes, have emerged as a class of promising two-dimensional (2D) materials since the discovery of monolayer $Ti_3C_2$ in 2011[1-7]. To date, more than 50 MXenes have been synthesized (MX structures, not counting solid solutions and distinct surface terminations), and over 100 distinct MXenes have been theoretically predicted based on combinations of M and X elements[8]. The chemical formula of MXenes is typically written as $M_{n+1}X_nT_x$, where M represents an early transition metal (lanthanides have been added recently[9, 10]), X represents carbon (C) or nitrogen (N), and $T_x$ represents surface functional groups such as oxygen (O), fluoride (F), chloride (Cl), hydrogen (H), hydroxyl (-OH), chalcogens, among others; $n$ is the layer number ($n$ = 1-4); and $x$ is the number of surface terminations per unit formula (typically $x$ = 2). Due to the large variety of transition metals (M) and surface functional groups ($T_x$), MXenes exhibit a huge (virtually infinite) compositional space and unique and highly tunable material properties, which provide them with a significant potential for various practical applications, such as catalysis[11, 12], superconductivity[6], energy storage[13], environmental remediation[14], electromagnetic devices[15], and many others[13, 16].

Experimentally, 2D layered MXenes are typically syntheiszed by selectively etching the "A" layers in their precursor MAX phases[17-19], $M_{n+1}AX_n$, followed by a delamination process using intercalation and mechanical agitation/sonication[20] or other techniques such as liquid exfoliation[21, 22]. It is widely recognized that MXenes can inherit similar crystal structures as those of layered MAX phases[23]. For instance, aberration-corrected scanning transition electron microscopy (AC-STEM) combined with refined X-ray (XRD) measurements[6] have revealed the structural relationships between the three-dimensional (3D) $Ti_3AlC_2$ MAX phases and their associated $Ti_3C_2T_x$ MXene, where the MXene inherits the hexagonal closed packed (HCP) arrangement of the MAX phases. One notable feature of this crystal structure is that the carbon atoms located between Ti layers are octahedrally bonded with six nearest-neighbor Ti elements, reminiscent of a face-centered cubic (FCC)-like local crystal structure[9, 24]. However, some experimental and computational studies have demonstrated that the X elements (C and/or N), located between M layers, can form a trigonal prismatic coordination environment with six surrounding M elements[25-36]. Notably, some MXene phases are more energetically and structurally stable in this form than the octahedral one[25-36]. These findings suggest that the stability of MXenes is influenced not only by their 3D precursor (e.g., MAX phases) but also by the local coordination arrangements of both X and M elements within the layered MXenes.

The study of phase stability in MXenes, specifically between octahedral and trigonal prismatic coordination, parallels prior research on transition metal dichalcogenides (TMDs)[37]. Using nomenclature similar to TMDs, a monolayer MXene with octahedral coordination can be termed as the 1T phase, while a structure with trigonal prismatic coordination can be referred to as the 1H phase (or 2H phase, considering termination elements in some studies[25]). Extensive studies have examined the stability of the 1T and 1H phases across a series of MXenes. For example, carbide-based MXenes with groups III and IV transition metals and nitride-based MXenes with group IV transition metals are typically stable in the 1T phases[26, 28, 32, 35], while both carbide- and nitride-based MXenes with group VI elements are more likely to form the 1H phases[26, 33, 35, 36, 38]. Similar phase stability trends have also been studied in layered transition metal borides, sulfides, and phosphides, with both 1T and 1H phases theoretically predicted in these materials[27, 39, 40]. Given the markedly different material properties exhibited by these phases, understanding their stability is a critical step toward developing novel MXenes with improved properties.

In this work, we systematically investigated the phase stability of MXenes with various local crystal structures, tracing the origins to their hierarchichal precursors, including the intermediate MAX phases and the parent bulk phases. We found that all MXenes, from monolayer to multiple layers ($n$ = 1-3), can be derived from four types of bulk phases, each characterized by unique atomic coordination environments[41, 42]. Using Ti- and Mo-based TMCs and TMNs as model systems, we carried out extensive density functional theory (DFT) calculations and demonstrated that these bulk phases have consistent energetic stability relative to their derived MXene counterparts and the intermediate MAX phases. Furthermore, high-throughput modeling suggests that this structural relationship extends across all transition metals in the periodic table, enabling the direct use of bulk phase information to predict MXene crystal structures without reliance on intermediate MAX phases (Fig. 1a). This study significantly expands the landscape of MXenes, at least doubling the number of possible structures.

## Results

### Nomenclature for MXene structures

Two types of coordinations are commonly observed in 3D bulk TMCs and TMNs, and they are respectively octahedral (denoted as O) and trigonal prismatic (denoted as P), where the former has six atoms that form bond angles of 90° around the center atom, while the latter has six atoms arranged in a form of triangular prism. Since both the M and X elements can exhibit these two types of coordination, the combination of them gives four types of bulk structures with various coordination environments (Fig. 1b) with a 1:1 stoichiometric ratio of M to X elements. The first bulk phase has an FCC, rock-salt crystal structure with a coordination sequence O-O respectively for M and X elements. This classical structure, with a space group of $Fm\bar{3}m$, is the most widely observed local crystal structure for the central layers inside

MXenes. For convenience, we denote this type of crystal structure as the B1 phase in this study. Additionally, two nickel arsenide (NiAs)-type hexagonal crystal structures[43] with both O and P coordination are illustrated in Fig. 1b. The first NiAs-type hexagonal structure, denoted as HX1a, has a coordination sequence of P-O, respectively, for M and X elements. The second NiAs-type hexagonal structure has an exchanged coordination sequence O-P for M and X elements, and we denote it as HX1b. These two hexagonal structures have a space group of $P6_3/mmc$, which is commonly observed in the MAX phases ($M_{n+1}AX_n$) or terminated MXenes ($M_{n+1}X_nT_x$), as the main group elements A or termination groups T can form such coordination arrangement with surrounding M elements. Finally, a tungsten carbide (WC)-type hexagonal crystal structure, with a space group of $P\bar{6}m2$[44], exhibits a P-P coordination sequence for M and X elements. We denoted this type of bulk phase as HX2; see Fig. 1b.

Each bulk phase can yield two variants of MXenes with distinct coordination sequences characterized by different atomic positions of the surface termination elements. Figs. 1c-e illustrate these bulk-derived MXenes, ranging from the thinnest $M_2XT_2$ to $M_4X_3T_2$. For $M_2XT_2$ (Fig. 1c), the B1-derived MXenes have two types of coordination sequence for M-X-M: O'-O-O' and P'-O-P', based on the termination element positions, where the prime symbol (') denotes the coordination of the outermost layer of M elements. These structures are referred to as T-1 and T-2, respectively (Fig. 1c). Owing to the octahedral coordination of the middle layer of X element, the HX1a-derived MXenes share the same coordination sequence as the B1-derived MXenes, *i.e.*, T-2 and T-1 MXenes (Fig. 1c). In contrast, HX1b- and HX2-derived MXenes, with prismatic coordination for the middle layer of X elements, resulting in two new MXenes with coordination sequences O'-P-O' and P'-P-P', referred to as H-1 and H-2 MXenes, respectively (Fig. 1c). Overall, four types of $M_2XT_2$-based MXenes with different coordination sequences have been derived from the four bulk phases.

Thicker MXenes, $M_3X_2T_2$, have a more complicated coordination environment than thinner variants. Specifically, the two B1-derived MXenes, T-1 and T-2, have two distinct coordination sequences for the M-X-M-X-M arrangement: O'-O-O-O' and P'-O-O-O-P' (Fig. 1d). Different from the thinnest cases, these two MXenes are not the same as HX1a-derived ones, as the central M layer in the latter configuration has prismatic coordination, leading to P'-O-P-O-P' and O'-O-P-O-O' coordination sequences (named as H1a-1 and H1a-2). Similarly, HX1b-derived MXenes exhibit O'-P-O-P-O' and P'-P-O-P-P' coordination sequences, named H1b-1 and H1b-2, respectively (Fig. 1d). In HX2-derived $M_3X_2T_2$ MXenes, both M and X elements adopt prismatic coordination, resulting in P'-P-P-P-P' and O'-P-P-P-O' coordination sequences, which are named as H2-1 and H2-2. Eight different $M_3X_2T_2$-based MXenes with various coordination sequences can be derived from the four bulk phases.

Similar to $M_3X_2T_2$, $M_4X_3T_2$ MXenes also have eight bulk-derived structures with different coordination sequences (Fig. 1e). Using the same nomenclature, the B1-derived MXenes are designated as T-1 and T-2; HX1a-derived MXenes as H1a-1 and H1a-2; HX1b-derived MXenes as H1b-1 and H1b-2; and HX2-derived MXenes as H2-1 and H2-2 (Fig. 1e).

In total, 20 MXenes (4+8+8) can be derived from the four bulk phases across all three groups of MXenes with different thicknesses. Since most experimentally reported MXenes exhibit B1-derived crystal structures (T-1 and T-2), we estimate that this study effectively doubles the number of currently known MXenes.

**Phase stability of MXenes**

With a clear understanding of the crystal structures of MXenes and their associated coordination sequence, we performed DFT calculations to analyze the energetic relationships between the bulk phases, MAX phases, and their derived MXenes. Due to existing experiments, we primarily selected Ti- and Mo-based TMCs and TMNs in this work and chose aluminum (Al) as the "A" layer in the MAX phases[4]. Table 1 summarizes the DFT-calculated phase stability of TiC-based bulk phases, MAX phases, and corresponding MXenes. These results indicate that the B1 phase with an O-O coordination is the most stable, suggesting that octahedral coordination is the most favorable for Ti and C elements. This finding agrees with many prior experimental and computational studies[4, 32].

Interestingly, both TiC-based MAXs and MXenes have similar trends of phase stability as bulk phases. For instance, B1-derived $Ti_2AlC$ (MAX) and $Ti_2CF_2$ (MXene) both exhibit the same order of phase stability as their corresponding bulk phases, reinforcing that octahedral coordination is beneficial to both Ti and C elements in MAX and MXenes. Even in the thicker MAX and MXenes, Table 1 also shows that B1-derived phases, T-1 or T-2, have the most stable structure among all MXenes. This prediction is in line with the trend of bulk phases and thus suggests that all TiC-based $Ti_{n+1}C_nF_2$ ($n$ = 1-3) MXenes synthesized from the MAX route by HF etching should end up with the B1-derived MXenes.

Similar phase stability trends were found between bulk phases and their associated MAX and MXenes phases for Mo-based carbides. According to DFT calculations, the most stable bulk phase of MoC is HX2 with P-P coordination, suggesting that both Mo and C preferably adopt trigonal prismatic coordination. This observation is consistent with prior studies[25, 28, 33]. By examining the energetic stability of MoC bulk-derived MAX phases and MXenes, we found that HX2-derived structures are generally the most stable, especially when $n \geq 2$. This finding indicates that MoC-based MXenes synthesized by the MAX route through HF etching should end up with HX2-derived structures, where P coordination dominates (Table 2). One exception was found for the thinnest case, $Mo_2CF_2$, where the most stable configuration in the $Mo_2AlC$ MAX phase is B1-derived T-1. This exception is probably because Al layers are more favorable in forming octahedral coordinates with the outermost Mo layers. Still, after etching the Al layers, the monolayer $Mo_2CF_2$ becomes more energetically favorable in a hexagonal structure.

Furthermore, we systematically performed DFT calculations to evaluate the structural stability of nitrides-based MAX phases and MXenes for all bulk-derived configurations. Our results consistently show that TiN- and MoN-based MAX and MXenes always exhibit a very similar order of phase stability compared to their bulk counterparts (Suppl. Tables S1-2). These findings further indicate that bulk phases can serve as an effective indicator to predict the stable structures of MXenes in terms of their coordination environment; we carried out DFT calculations for 240 MXenes structures for Ti- and Mo-based carbide and nitride MXenes with three different termination groups (oxygen termination, fluoride termination, and non-termination). All DFT-calculated energies and lattice parameters of these MXenes are documented in Suppl. Tables S3-16.

**Lattice-dynamical stability of MXenes**

Building on the prior analysis, we conclude that MXenes can be rationally designed based on the stability of their bulk counterparts. However, confirming their lattice dynamical stability is crucial to ensure they represent local minima on the potential energy surface and are viable for experimental synthesis. Thus, we performed density functional perturbation theory (DFPT)-based phonon calculations to evaluate the lattice-dynamical stability of all MXenes studied in this work. Based on the calculated phonon spectra (Suppl. Figs. S1-30), we identified the stable MXenes of all Mo- and Ti-based MXenes with three termination groups: non-termination, F-termination, and O-termination.

An analysis of the stable MXenes for non-termination cases (Figs. 2a-c) reveals that 76 out of 80 (~95%) MXenes across all three thicknesses considered ($n$ = 1-3) are lattice dynamically stable without any imaginary frequency, except a few cases in the $Mo_3N_2$ and $Ti_4N_3$ families. These results suggest that the novel MXenes studied in this work, especially those derived from hexagonal bulk phases such as HX1a-, HX1b-, and HX2-derived MXenes, are lattice dynamically stable and feasible to be synthesized under vacuum and observed in experiments.

Phonon calculations were also conducted for F- and O-terminated MXenes. Using TiC-based MXenes as one example, some representative phonon spectra of hexagonal-derived MXenes are plotted in Fig. 2d-f. The phonon spectra reveal that most F-terminated TiC MXenes do not exhibit any imaginary frequency, suggesting lattice dynamical stability. Yet, O-terminated TiC MXenes consistently exhibit large imaginary frequencies (Fig. 2d-f), indicating that oxygen atoms are less efficient in stabilizing MXene structures. As summarized in Fig. 2a-c, F-terminated MXenes have more stable configurations than O-terminated cases. For instance, 58 out of 80 (72.5%) F-terminated MXenes are dynamically stable compared to only 45 out of 80 (56.3%) O-terminated MXenes. This finding suggests F atoms are more effective than O atoms in stabilizing layered structures of MXene, implying that future experiments should adopt hydrogen fluoride to etch the MAX phases to achieve these stable MXenes.

**A comprehensive search for MXene phases**

Although DFT calculations have demonstrated that bulk phase stability can effectively predict the stability of MXenes for Ti- and Mo-based systems, the bulk phase stability for other transition metals remains largely unexplored. Therefore, we carried out high-throughput DFT calculations across the periodic tables for transitional metals to calculate the phase stability of their bulk phases based on their ground-state energies. By comparing the relative stabilities of four bulk phases, we identified 13 different ranking orders, which are illustrated by distinct colors across the periodic table in Fig. 3a-b.

Fig. 3a shows the phase stability order of all TMCs. It is evident that group III, IV, and V elements predominantly appear in dark purple, indicating that the most stable MXenes based on these metals should be B1-derived phases. Such a prediction agrees with some prior studies, which have found that Sc-, Ti-, Zr-, Nb-, and Hf-based monolayer carbide MXenes are most stable in an octahedral coordination environment[26]. On the contrary, the yellow regions corresponding to group VI to VIII metals suggest that their favorable bulk phases are HX2, implying that layered MXenes based on these elements are likely to form HX2-based structures.

To verify this trend, we performed DFT calculations to assess the stability of MXenes for two representative transitional metals, *i.e.*, Hf and Re elements, which favor B1- and HX2-bulk phases. As shown in Fig. 3c, the most stable HfC-based MXenes across three thicknesses (n=1-3) are consistently B1-derived T-1 or T-2, but the most stable ReC-based MXenes are always HX2-derived or H1b-derived. These results further confirmed our prior findings that bulk phase stability can be exploited to reliably predict the structural stability of MXenes. It is also worth noting that the most stable bulk phase of group IX to XII metals are either HX1a or HX1b phases, meaning that their associated MXenes will likely adopt these two structures.

Fig. 3b presents the order of phase stability for all TMNs. Interestingly, the color map of this phase stability has one group leftward shift compared to TMCs, which can be ascribed to nitrogen having one extra electron than carbon atoms. The most stable B1 phases of TMNs are observed for groups III, IV, XII, and some group XI metals (dark purple colors). Meanwhile, the HX2 stable bulk phase is mainly from group V metals as well as Cr and Tc. However, there is no clear boundary distinguishing the stability of HX2 and HX1 phases between group VI and X elements (Fig. 3b).

To validate these trends, we again used Hf and Re as representatives to evaluate the phase stability of their MXenes across three layers (*n* = 1-3). As shown in Fig. 3d, HfN-based MXenes consistently exhibit B1-derived (T-2) structures, while ReN-based MXenes tend to have H1b- or HX2-derived structures. Notably, the most favorable B1 phase of HfN suggests that their derived MXenes are highly favorable in octahedral environments, which agrees well with prior observations[28]. Therefore, our calculations once again confirm that bulk phase stability can be used as an indicator to predict the phase stability of layered MXenes.

**Discussion**

MXenes are currently synthesized through a top-down approach, starting with their bulk precursor MAX phases, which are then exfoliated into layered 2D phases. While the resulting layered MXenes are typically considered to maintain the same crystal structure as their parent MAX phases, this is not always the case. Reviewing the coordination environment shows that all these 2D phases can be derived from four distinct bulk phases. Subsequent DFT calculations indicate that the structural stability of MXene is always consistent with the phase stability of their bulk counterparts. These findings suggest a design strategy for 2D materials, where the bulk phase can be used as an important indicator to predict the structures and stability of derived MXenes.

Fig. 3e summarizes all reported MXene structures from experiments and modeling efforts since the first discovery of MXene in 2011. The data reveal that the reported MXenes primarily exhibit B1-derived or HX1a-derived structures dominated by the O- or P-type coordination environments, respectively. By extending the analysis to four bulk phases in this study, we show that the internal bulk crystal structure of MXenes, particularly in multiple-layer cases ($n \geq 2$), can have alternating O- and P-coordination sequences. Recent studies have further demonstrated that MXenes can behave as metastable bulk phases at elevated temperatures, as evidenced by the phase transition of the ~1 nm-thick $Ti_3C_2T_x$ from hexagonal crystal to the B1 crystal structure[45-47]. Thus, it is reasonable to anticipate that these thick $Ti_3C_2T_x$ can alternate between O and P coordination during the phase transition. Meanwhile, in the presence of external stimuli, MXene may undergo phase-like transformation among the various 2D phases by alternating the coordination sequence, a phenomenon that has been observed in TMD and MAX phases[48, 49]. Therefore, revisiting the crystal structures and their coordination environment of all MXenes is particularly important to understand phase transition behaviors in MXenes.

Finally, this study establishes a structural library of MXenes that extends beyond layered TMC and TMNs to encompass other transition metal-based 2D materials. For instance, layered transition metal borides (MBenes)[40, 50], sulfides (MSenes)[39, 51], and phosphides (MPenes)[52, 53] have been investigated experimentally and computationally. However, most prior studies primarily focused on the 1T and 2H monolayer phases derived from TMD research. Future studies of these 2D materials can expand to include other bulk-derived 2D phases, which remain largely unexplored. The panoramic view of 2D phases lays a solid foundation for understanding phase stability and their structural relationships across a broad range of 2D materials.

**Methods**

**Density functional theory (DFT) calculations**

All density functional theory (DFT) calculations were conducted using the Vienna ab initio Simulation Package (VASP, version 5.4.4)[54-56] and the projector-augmented wave (PAW) method[57, 58]. The PBE-D2 functional[59] with vdW correction was adopted for the structural optimizations of all MXene phases, where all atoms were relaxed until the Hellmann-Feynman forces were smaller than $10^{-2}$ eV/Å. Following the structural optimizations, we performed static calculations to calculate the charge density and electronic band structures using the semi-local Perdew-Burke-Ernzerhof (PBE) exchange-correlation functional[60]. The lattice parameters along all directions were fully optimized for the bulk and MAX phases, and only Al-containing MAX phases are considered in this study. However, in the case of 2D materials, the lattice parameters were only allowed to relax along the *x*- and *y*-directions (parallel to the 2D materials). Meanwhile, the lattice parameter along the z direction (perpendicular to the 2D materials) was fixed at 20 Å to avoid interactions from adjacent periodic boundaries for all kinds of calculations. According to the convergence test, the Brillouin-zone integrations were performed using a $\Gamma$-centered 14×14×1 *k*-point grid for all MXene and MAX phases and a 10×10×10 grid for all cubic unit cells. The kinetic energy cutoff for plane waves was set to 700 eV, the convergence criterion for electronic self-consistency was set to $10^{-6}$ eV, and VASP's "accurate" precision setting was adopted to avoid wrap-around errors. Spin polarization was considered for all DFT calculations due to the magnetization observed in some MXenes. All crystal structures were visualized by VESTA software[61].

**Phonon calculations**
Based on the DFT-optimized structures, we calculated the phonon spectra and vibrational density of states (DOS) for all MXenes using the Phonopy code[62, 63]. The density functional theory perturbation theory (DFPT) method was adopted to calculate the force constants. Here, the energy cutoff, energy convergence, and force convergence were the same as those of structural relaxations. We tested the 1×1×1, 2×2×1, and 3×3×1 supercells of MXene phases to eliminate the imaginary phonon frequencies. We found that 3×3×1 supercell is the sufficient to achieve clean phonon spectra; we, therefore, adopted this configuration for all phonon calculations. The phonon spectra were sampled in the reciprocal space using primitive-based high symmetry *k*-points $\Gamma$ (0, 0, 0), *K* (1/3, 1/3, 0), and *M* (0.5, 0, 0).

**Author contributions**
C.H. and J.H. initiated this study and supervised all aspects of this work. C.H. and J.H. performed initial DFT calculations, and N.O. and O.C. performed additional DFT calculations. T.D., B.S., P.K., and P.G. provided detailed feedback on the simulation method. All authors wrote the manuscript together and contributed to the revision of this manuscript.

**Acknowledgments**
N.O., O.C., and C.H. acknowledge the support of DOE Award DE-SC0025431. This research used resources of the National Energy Research Scientific Computing Center (NERSC), a DOE Office of Science User Facility supported by the Office of Science of the U.S. Department of Energy under Contract No. DE-AC02-05CH11231 using NERSC award BES-ERCAP0031213. A portion of the calculations used the resources of the Compute and Data Environment for Science (CADES) at ORNL and of the NERSC, which are supported by the Office of Science of the U.S. DOE under contract nos. DE-AC05-00OR22750 and DE-AC02-05CH11231, respectively. Some work was performed at the Center for Nanophase Materials Sciences, a U.S. DOE Office of Science User Facility operated at Oak Ridge National Laboratory.

## Competing interests

The authors declare no competing interests.

## Additional information

**Supplementary information**: The authors prepared 16 Suppl. Tables and 30 Suppl. Figs. in the Supplementary Information.

semiconductor transition. *Turkish Journal of Physics* **2019**, *43* (5).

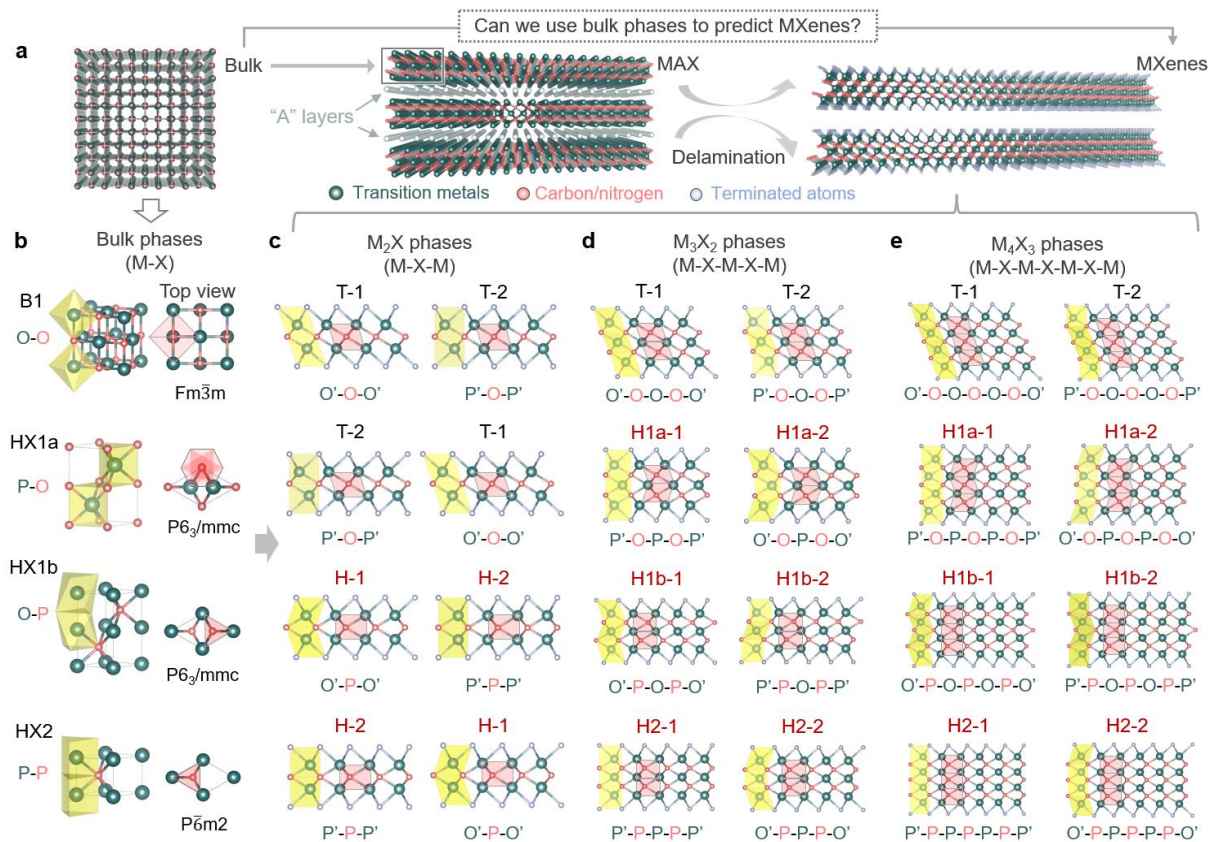

**Fig. 1. A panoramic view of 2D MXene structures in transition metals carbides (TMCs) or nitrides (TMNs).** **a** Schematic diagram of synthesizing the MXene structures from their hierarchical precursors, including the intermediate MAX phases and their parent bulk phases. Each TMC or TMN can have four types of bulk phases, and their crystal structures and associated top views are illustrated in panel **b**. These structures include one face-center-cubic (FCC) crystal structure (denoted as B1), two NiAs-type hexagonal structures (denoted as HX1a and HX1b), and one WC-type hexagonal crystal structure (denoted as HX2). The coordination number sequence of M-X (M = transition metals, X = C or N) was labeled for each structure. The octahedral and prismatic coordination were labeled as O and P, and the yellow and pink polyhedral, respectively, represent the coordination of M and X elements. **c** Crystal structures and associated coordination sequence (M-X-M) of four $M_2X$-type MXenes derived from each bulk phase. **d** Crystal structures and associated coordination sequence (M-X-M-X-M) of eight different $M_3X_2$-based MXenes derived from each bulk phase. **e** Crystal structures and associated coordination sequence (M-X-M-X-M-X-M) of eight different $M_4X_3$-based MXenes derived from each bulk phase. The B1-derived MXenes are commonly reported MXene phases observed in experiments. Based on the additional hexagonal bulk phases, we can derive and expand the candidates of the 2D MXenes family to 20 different structures. The symbol ' denotes the coordination of the outermost layer of M or X.

**Table 1. DFT-calculated phase stability and coordination environment for TiC-based bulk materials, MAX phases, and MXene phases**. The thinnest MAX phase (Ti$_2$CAl) is derived from four bulk structures, while thicker MAX phases (Ti$_3$C$_2$Al and Ti$_4$C$_3$Al) exhibit eight derived phases, separated into two major groups (1 and 2). The 2D MXenes are tabulated similarly to MAX phases, and only F-terminated MXenes are summarized in this table. The order of phase stability is based on the ranking of DFT-calculated ground-state energies. The most stable coordination is listed for each group.

| Phase | Materials system | Coordination | Order of phase stability | Most stable coordination |
|---|---|---|---|---|
| Bulk | TiC | Ti-C | B1 < HX1a < HX1b < HX2 | O-O |
| MAX | Ti$_2$CAl | Ti-C-Ti | T-1 < T-2 < H-1< H-2 | O'-O-O' |
| | Ti$_3$C$_2$Al-1 | Ti-C-Ti-C-Ti | T-1 < H1a-1 < H1b-1 < H2-1 | O'-O-O-O-O' |
| | Ti$_3$C$_2$Al-2 | Ti-C-Ti-C-Ti | H1a-2 < T-2 < H1b-2 < H2-2 | O'-O-P-O-O' |
| | Ti$_4$C$_3$Al-1 | Ti-C-Ti-C-Ti-C-Ti | T-1 < H1a-1 < H1b-1 < H2-1 | O'-O-O-O-O-O-O' |
| | Ti$_4$C$_3$Al-2 | Ti-C-Ti-C-Ti-C-Ti | T-2 < H1a-2 < H1b-2 < H2-2 | P'-O-O-O-O-O-P' |
| MXene | Ti$_2$CF$_2$ | Ti-C-Ti | T-1 < T-2 < H-1 < H-2 | O'-O-O' |
| | Ti$_3$C$_2$F$_2$-1 | Ti-C-Ti-C-Ti | T-1 < H1a-1 < H1b-1 < H2-1 | O'-O-O-O-O' |
| | Ti$_3$C$_2$F$_2$-2 | Ti-C-Ti-C-Ti | H1a-2 < T-1 < H1b-2 < H2-2 | O'-O-P-O-O' |
| | Ti$_4$C$_3$F$_2$-1 | Ti-C-Ti-C-Ti-C-Ti | T-1 < T-2 < H-1 < H-2 | O'-O-O-O-O-O-O' |
| | Ti$_4$C$_3$F$_2$-2 | Ti-C-Ti-C-Ti-C-Ti | H1a-2 < T-1 < H1b-2 < H2-2 | O'-O-P-O-P-O-O' |

**Table 2. DFT-calculated phase stability and coordination environment for MoC-based bulk materials, MAX phases, and MXene phases**. The thinnest MAX phase (Mo$_2$CAl) is derived from four bulk structures, while thicker MAX phases (Mo$_3$C$_2$Al and Mo$_4$C$_3$Al) exhibit eight derived phases, separated into two major groups (1 and 2). The 2D MXenes are tabulated similarly to MAX phases, and only F-terminated MXenes are summarized in this table. The order of phase stability is based on the ranking of DFT-calculated ground-state energies. The most stable coordination is listed for each group.

| Phase | Materials system | Coordination | Order of phase stability | Most stable coordination |
|-------|------------------|--------------|--------------------------|--------------------------|
| Bulk | MoC | Mo-C | HX2 < HX1a < HX1b < B1 | P-P |
| MAX | Mo$_2$CAl | Mo-C-Mo | T-1 < T-2 < H-1< H-2 | O'-O-O' |
| | Mo$_3$C$_2$Al-1 | Mo-C-Mo-C-Mo | H2-1 < H1b-1 < H1a-1 < T-1 | P'-P-P-P-P' |
| | Mo$_3$C$_2$Al-2 | Mo-C-Mo-C-Mo | H2-2 < H1a-2 < H1b-2 < T-2 | O'-P-P-P-O' |
| | Mo$_4$C$_3$Al-1 | Mo-C-Mo-C-Mo-C-Mo | H2-1 < H1a-1 < H1b-1 < T-1 | P'-P-P-P-P-P-P' |
| | Mo$_4$C$_3$Al-2 | Mo-C-Mo-C-Mo-C-Mo | H2-2 < H1a-2 < H1b-2 < T-2 | O'-P-P-P-P-P-O' |
| MXene | Mo$_2$CF$_2$ | Mo-C-Mo | H-1 < H-2 < T-2 < T-1 | O'-P-O' |
| | Mo$_3$C$_2$F$_2$-1 | Mo-C-Mo-C-Mo | H1b-1 < H2-1 < T-1 < H1a-1 | O'-P-O-P-O' |
| | Mo$_3$C$_2$F$_2$-2 | Mo-C-Mo-C-Mo | H2-2 < H1b-2 < T-2 < H1a-2 | O'-P-P-P-O' |
| | Mo$_4$C$_3$F$_2$-1 | Mo-C-Mo-C-Mo-C-Mo | H2-1 < H1b-1 < T-1 < H1a-1 | P'-P-P-P-P-P-P' |
| | Mo$_4$C$_3$F$_2$-2 | Mo-C-Mo-C-Mo-C-Mo | H2-2 < H1b-2 < T-2 < H1a-2 | O'-P-P-P-P-P-O' |

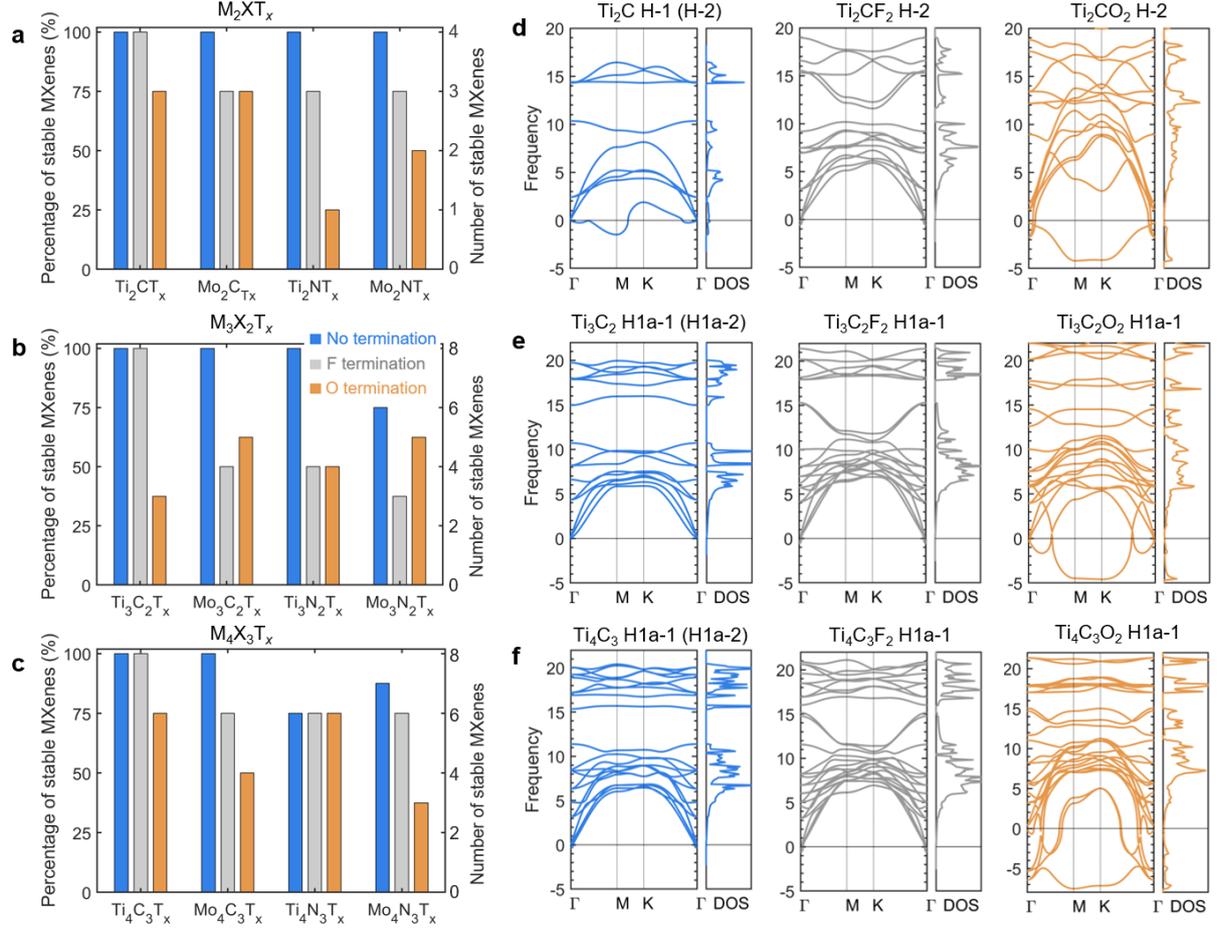

**Fig. 2. Lattice-dynamical stability of 2D MXenes**. **a-c** Percentage of lattice-dynamically stable 2D MXenes among all Mo- and Ti-based 2D MXenes with and without termination atoms for $M_2XT_x$, $M_3X_2T_x$, and $M_3X_2T_x$. Configurations with no imaginary frequencies or only minor ones are considered stable. **d** DFT-calculated phonon spectra and vibrational density of states (vDOS) of 2D hexagonal-derived $Ti_2C$ MXenes without termination, and with F and O terminations. **e-f** Phonon spectra and vDOS of 2D H1a-derived $Ti_3C_2$ and $Ti_4C_3$ MXenes under the same termination conditions.

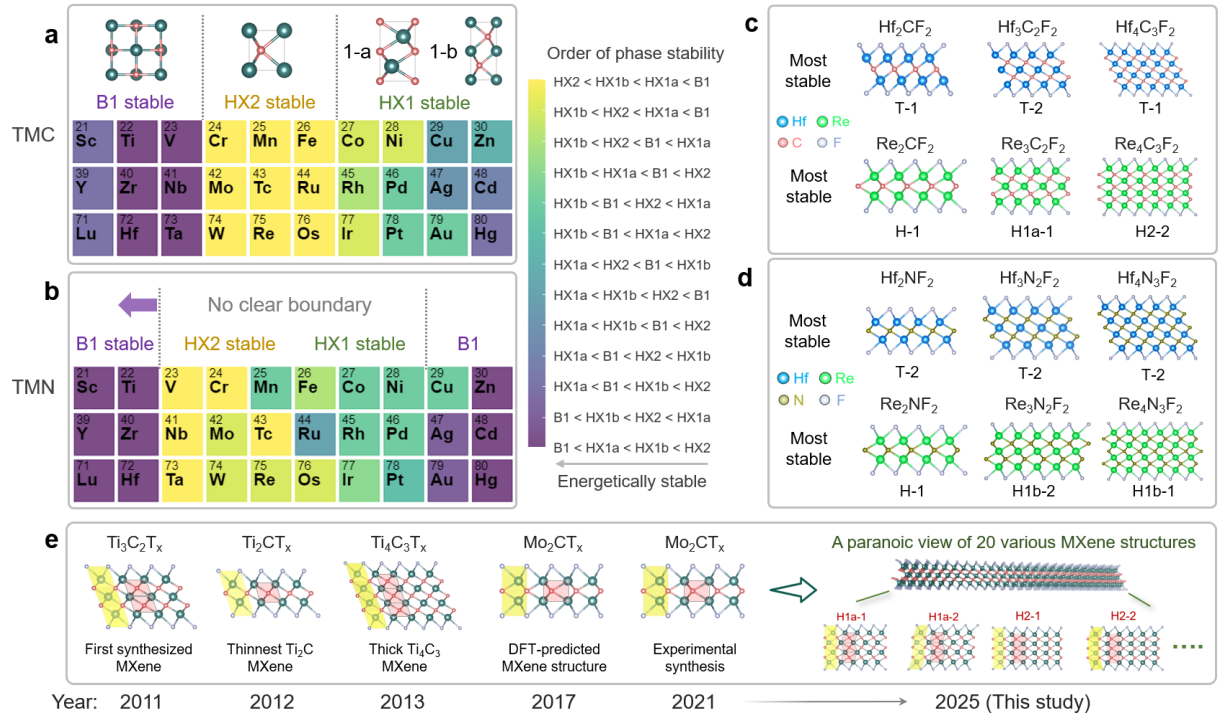

**Fig. 3. Comprehensive search for MXene phases based on bulk phase stability**. **a** Phase stability ordering of four bulk phases for all transitional metal carbides (TMC), ranked by DFT-calculated ground-state energies. Lower order suggests a more energetically stable phase. **b** Phase stability ordering of four bulk phases for all transitional metal nitrides (TMN). **c-d** Verification of phase stability of two representative 2D MXenes systems, Hf- and Re-based TMC and TMNs. **e** Chronological review of the existing MXene crystal structures reported by prior experiments and simulations. This study expands the family of 2D MXenes to 20 crystal structures based on the coordination environment, approximately doubling the number of MXenes reported so far.

**A Panoramic View of MXenes via a New Design Strategy**

Noah Oyeniran[a], Oyshee Chowdhury[a], Chongze Hu[a*], Traian Dumitrica[b], Panchapakesan Ganesh[c], Jacek Jakowski[d], Zhongfang Chen[e], Raymond R. Unocic[f], Michael Naguib[g], Vincent Meunier[h], Yury Gogotsi[i], Paul R. C. Kent[d], Bobby G. Sumpter[c], Jingsong Huang[c*]

[a] *Department of Aerospace Engineering and Mechanics, The University of Alabama, Tuscaloosa, Alabama 35487, USA*

[b] *Department of Mechanical Engineering, University of Minnesota Twin Cities, Minneapolis, Minnesota 55455, USA*

[c] *Center of Nanophase Materials Sciences, Oak Ridge National Laboratory, Oak Ridge, Tennessee 37831, USA*

[d] *Computational Sciences and Engineering Division, Oak Ridge National Laboratory, Oak Ridge, Tennessee 37831, USA*

[e] *Department of Chemistry, University of Puerto Rico, Rio Piedras, San Juan, Puerto Rico 00931, USA*

[f] *Department of Materials Science and Engineering, North Carolina State University, Raleigh, North Caroline 27695, USA*

[g] *Department of Physics and Engineering Physics, Tulane University, New Orleans, Louisiana 70118, USA*

[h] *Department of Engineering Science and Mechanics, The Pennsylvania State University, University Park, Pennsylvania 16801, USA*

[i] *Department of Materials Science and Engineering, and A.J. Drexel Nanomaterials Institute, Drexel University, Philadelphia, Pennsylvania 19104, USA*

*Corresponding authors
Email address: hucz@ua.edu (Chongze Hu) and huangj3@ornl.gov (Jingsong Huang)

# Table of Contents

**Table S1**. **DFT-calculated phase stability and coordination environment for TiN-based bulk materials, MAX phases, and MXene phases**. The thinnest MAX phase (Ti$_2$NAl) is derived from four bulk structures, while thicker MAX phases (Ti$_3$N$_2$Al and Ti$_4$N$_3$Al) exhibit eight derived phases, separated into two major groups (1 and 2). The 2D MXenes are tabulated similarly to MAX phases, and only F-terminated MXenes are summarized in this table. The order of phase stability is based on the ranking of DFT-calculated ground-state energies. The most stable coordination is listed for each group.

| Phase | Materials system | Coordination | Order of phase stability | Most stable coordination |
|---|---|---|---|---|
| Bulk | TiN | Ti-N | B1 < HX1a < HX1b < HX2 | O-O |
| MAX | Ti$_2$NAl | Ti-N-Ti | T-1 < T-2 < H-1< H-2 | O'-O-O' |
| | Ti$_3$N$_2$Al-1 | Ti-N-Ti-N-Ti | T-1 < H1a-1 < H1b-1 < H2-1 | O'-O-O-O-O' |
| | Ti$_3$N$_2$Al-2 | Ti-N-Ti-N-Ti | T-2 < H1a-2 < H1b-2 < H2-2 | P'-O-O-O-P' |
| | Ti$_4$N$_3$Al-1 | Ti-N-Ti-N-Ti-N-Ti | T-1 < H1a-1 < H1b-1 < H2-1 | O'-O-O-O-O-O-O' |
| | Ti$_4$N$_3$Al-2 | Ti-N-Ti-N-Ti-N-Ti | T-2 < H1a-2 < H1b-2 < H2-2 | P'-O-O-O-O-O-P' |
| MXene | Ti$_2$NF$_2$ | Ti-N-Ti | T-1 < T-2 < H-2 < H-1 | O'-O-O' |
| | Ti$_3$N$_2$F$_2$-1 | Ti-N-Ti-N-Ti | T-1 < H1a-1 < H2-1 < H1b-1 | O'-O-O-O-O' |
| | Ti$_3$N$_2$F$_2$-2 | Ti-N-Ti-N-Ti | T-2 < H1a-2 < H2-2 < H1b-2 | P'-O-O-O-P' |
| | Ti$_4$N$_3$F$_2$-1 | Ti-N-Ti-N-Ti-N-Ti | T-1 < H1a-1 < H1b-1 < H2-1 | O'-O-O-O-O-O-O' |
| | Ti$_4$N$_3$F$_2$-2 | Ti-N-Ti-N-Ti-N-Ti | T-2 < H1a-2 < H1b-2 < H2-2 | P'-O-O-O-O-O-P' |

**Table S2**. **DFT-calculated phase stability and coordination environment for MoN-based bulk materials, MAX phases, and MXene phases**. The thinnest MAX phase ($Mo_2NAl$) is derived from four bulk structures, while thicker MAX phases ($Mo_3N_2Al$ and $Mo_4N_3Al$) exhibit eight derived phases, separated into two major groups (1 and 2). The 2D MXenes are tabulated similarly to MAX phases, and only F-terminated MXenes are summarized in this table. The order of phase stability is based on the ranking of DFT-calculated ground-state energies. The most stable coordination is listed for each group.

| Phase | Materials system | Coordination | Order of phase stability | Most stable coordination |
|---|---|---|---|---|
| Bulk | MoN | Mo-N | HX1b < HX2 < HX1a < B1 | O-P |
| MAX | $Mo_2NAl$ | Mo-N-Mo | H-1 < H-2 < T-2 < T-1 | O'-P-O' |
| | $Mo_3N_2Al$-1 | Mo-N-Mo-N-Mo | H1b-1 < H2-1 < H1a-1 < T-1 | O'-P-O-P-O' |
| | $Mo_3N_2Al$-2 | Mo-N-Mo-N-Mo | H1b-2 < H2-2 < H1a-2 < T-2 | P'-P-O-P-P' |
| | $Mo_4N_3Al$-1 | Mo-N-Mo-N-Mo-N-Mo | H1b-1 < H2-1 < H1a-1 < T-1 | O'-P-O-P-O-P-O' |
| | $Mo_4N_3Al$-2 | Mo-N-Mo-N-Mo-N-Mo | H1b-2 < H2-2 < H1a-2 < T-2 | P'-P-O-P-O-P-P' |
| MXene | $Mo_2NF_2$ | Mo-N-Mo | H-1 < T-1 < H-2 < T-2 | O'-P-O' |
| | $Mo_3N_2F_2$-1 | Mo-N-Mo-N-Mo | H1b-1 < H2-1 < T-1 < H1a-1 | O'-P-O-P-O' |
| | $Mo_3N_2F_2$-2 | Mo-N-Mo-N-Mo | H2-2 < H1b-2 < H1a-2 < T-2 | O'-P-P-P-O' |
| | $Mo_4N_3F_2$-1 | Mo-N-Mo-N-Mo-N-Mo | H1b-1 < T-1 < H2-1 < H1a-1 | O'-P-O-P-O-P-O' |
| | $Mo_4N_3F_2$-2 | Mo-N-Mo-N-Mo-N-Mo | H2-2 < H1b-2 < H1a-2 < T-2 | O'-P-P-P-P-P-O' |

**Table S3**. DFT-calculated energy difference (ΔE) and DFT-optimized lattice constants (*a*) of all non-terminated TiC-based 2D MXenes ($Ti_{n+1}C_n$, where *n* = 1-3) using PBE functional and PBE-D2 with vdW corrections. The energy difference is determined by: $\Delta E = E_{MX} - \min(E_{MX})$, where $E_{MX}$ is the ground state energy of each MXene structure and $\min(E_{MX})$ is the energy of the most stable MXene within the same group.

| | PBE-D2 ΔE (eV/atom) | PBE ΔE (eV/atom) | PBE-D2 *a* (Å) | PBE *a* (Å) |
|---|---|---|---|---|
| Ti₂C (no termination) | | | | |
| T-1 | 0.000 | 0.000 | 3.04 | 3.04 |
| T-2 | 0.000 | 0.000 | 3.04 | 3.04 |
| H-1 | 0.426 | 0.424 | 3.02 | 3.01 |
| H-2 | 0.426 | 0.424 | 3.02 | 3.01 |
| Ti₃C₂ (no termination) | | | | |
| T-1 | 0.000 | 0.000 | 3.10 | 3.10 |
| T-2 | 0.000 | 0.000 | 3.10 | 3.10 |
| H1a-1 | 0.034 | 0.030 | 3.09 | 3.09 |
| H1a-2 | 0.034 | 0.030 | 3.09 | 3.09 |
| H1b-1 | 0.477 | 0.464 | 3.06 | 3.06 |
| H1b-2 | 0.477 | 0.464 | 3.06 | 3.07 |
| H2-1 | 0.521 | 0.505 | 3.06 | 3.07 |
| H2-2 | 0.521 | 0.505 | 3.07 | 3.07 |
| Ti₄C₃ (no termination) | | | | |
| T-1 | 0.000 | 0.000 | 3.09 | 3.09 |
| T-2 | 0.000 | 0.000 | 3.09 | 3.09 |
| H1a-1 | 0.036 | 0.028 | 3.10 | 3.11 |
| H1a-2 | 0.036 | 0.028 | 3.10 | 3.11 |
| H1b-1 | 0.488 | 0.466 | 3.06 | 3.06 |
| H1b-2 | 0.488 | 0.466 | 3.06 | 3.06 |
| H2-1 | 0.555 | 0.529 | 3.07 | 3.08 |
| H2-2 | 0.555 | 0.529 | 3.07 | 3.08 |

**Table S4**. DFT-calculated energy difference ($\Delta E$) and DFT-optimized lattice constants ($a$) of all oxygen-terminated TiC-based 2D MXenes ($Ti_{n+1}C_nO_2$, where $n$ = 1-3) using PBE functional and PBE-D2 with vdW corrections. The energy difference is determined by: $\Delta E = E_{MX} - \min(E_{MX})$, where $E_{MX}$ is the ground state energy of each MXene structure and $\min(E_{MX})$ is the energy of the most stable MXene within the same group.

| Ti₂CO₂ (oxygen termination) | | | | |
|---|---|---|---|---|
| | PBE-D2 $\Delta E$ (eV/atom) | PBE $\Delta E$ (eV/atom) | PBE-D2 $a$ (Å) | PBE $a$ (Å) |
| T-1 | 0.000 | 0.000 | 3.03 | 3.03 |
| T-2 | 0.374 | 0.366 | 2.96 | 2.96 |
| H-1 | 0.087 | 0.072 | 3.04 | 3.05 |
| H-2 | 0.443 | 0.420 | 2.96 | 2.97 |
| Ti₃C₂O₂ (oxygen termination) | | | | |
| | PBE-D2 $\Delta E$ (eV/atom) | PBE $\Delta E$ (eV/atom) | PBE-D2 $a$ (Å) | PBE $a$ (Å) |
| T-1 | 0.000 | 0.000 | 3.04 | 3.04 |
| T-2 | 0.219 | 0.213 | 3.01 | 3.02 |
| H1a-1 | 0.298 | 0.290 | 3.15 | 3.15 |
| H1a-2 | 0.091 | 0.087 | 3.03 | 3.04 |
| H1b-1 | 0.186 | 0.165 | 3.04 | 3.05 |
| H1b-2 | 0.411 | 0.384 | 3.02 | 3.02 |
| H2-1 | 0.521 | 0.490 | 2.99 | 2.99 |
| H2-2 | 0.288 | 0.263 | 3.04 | 3.04 |
| Ti₄C₃O₂ (oxygen termination) | | | | |
| | PBE-D2 $\Delta E$ (eV/atom) | PBE $\Delta E$ (eV/atom) | PBE-D2 $a$ (Å) | PBE $a$ (Å) |
| T-1 | 0 | 0.000 | 3.04 | 3.04 |
| T-2 | 0.174 | 0.170 | 3.02 | 3.03 |
| H1a-1 | 0.288 | 0.278 | 3.10 | 3.11 |
| H1a-2 | 0.123 | 0.118 | 3.05 | 3.05 |
| H1b-1 | 0.251 | 0.225 | 3.04 | 3.05 |
| H1b-2 | 0.425 | 0.396 | 3.02 | 3.03 |
| H2-1 | 0.571 | 0.536 | 3.00 | 3.00 |
| H2-2 | 0.387 | 0.356 | 3.04 | 3.04 |

**Table S5**. DFT-calculated energy difference ($\Delta E$) and DFT-optimized lattice constants ($a$) of all fluoride-terminated TiC-based 2D MXenes ($Ti_{n+1}C_nF_2$, where $n$ = 1-3) using PBE functional and PBE-D2 with vdW corrections. The energy difference is determined by: $\Delta E = E_{MX} - \min(E_{MX})$, where $E_{MX}$ is the ground state energy of each MXene structure and $\min(E_{MX})$ is the energy of the most stable MXene within the same group.

| | | | | |
|---|---|---|---|---|
| Ti₂CF₂ (fluoride termination) | | | | |
| | PBE-D2 $\Delta E$ (eV/atom) | PBE $\Delta E$ (eV/atom) | PBE-D2 $a$ (Å) | PBE $a$ (Å) |
| T-1 | 0.000 | 0.000 | 3.05 | 3.06 |
| T-2 | 0.117 | 0.112 | 2.97 | 2.98 |
| H-1 | 0.213 | 0.211 | 3.07 | 3.08 |
| H-2 | 0.337 | 0.326 | 2.95 | 2.95 |
| Ti₃C₂F₂ (fluoride termination) | | | | |
| | PBE-D2 $\Delta E$ (eV/atom) | PBE $\Delta E$ (eV/atom) | PBE-D2 $a$ (Å) | PBE $a$ (Å) |
| T-1 | 0.000 | 0.000 | 3.07 | 3.08 |
| T-2 | 0.110 | 0.106 | 3.02 | 3.02 |
| H1a-1 | 0.145 | 0.138 | 3.01 | 3.02 |
| H1a-2 | 0.021 | 0.018 | 3.07 | 3.07 |
| H1b-1 | 0.293 | 0.280 | 3.06 | 3.06 |
| H1b-2 | 0.386 | 0.368 | 3.00 | 3.00 |
| H2-1 | 0.451 | 0.431 | 2.98 | 2.99 |
| H2-2 | 0.332 | 0.317 | 3.06 | 3.07 |
| Ti₄C₃F₂ (fluoride termination) | | | | |
| | PBE-D2 $\Delta E$ (eV/atom) | PBE $\Delta E$ (eV/atom) | PBE-D2 $a$ (Å) | PBE $a$ (Å) |
| T-1 | 0.000 | 0.000 | 3.08 | 3.08 |
| T-2 | 0.085 | 0.082 | 3.03 | 3.03 |
| H1a-1 | 0.148 | 0.140 | 3.03 | 3.04 |
| H1a-2 | 0.039 | 0.034 | 3.08 | 3.08 |
| H1b-1 | 0.328 | 0.309 | 3.05 | 3.06 |
| H1b-2 | 0.407 | 0.384 | 3.01 | 3.01 |
| H2-1 | 0.502 | 0.476 | 3.00 | 3.00 |
| H2-2 | 0.402 | 0.380 | 3.05 | 3.06 |

**Table S6**. DFT-calculated energy difference ($\Delta E$) and DFT-optimized lattice constants ($a$) of all non-terminated MoC-based 2D MXenes ($Mo_{n+1}C_n$, where $n$ = 1-3) using PBE functional and PBE-D2 with vdW corrections. The energy difference is determined by: $\Delta E = E_{MX} - \min(E_{MX})$, where $E_{MX}$ is the ground state energy of each MXene structure and $\min(E_{MX})$ is the energy of the most stable MXene within the same group.

| Mo₂C (no termination) | | | | |
|---|---|---|---|---|
| | PBE-D2 $\Delta E$ (eV/atom) | PBE $\Delta E$ (eV/atom) | PBE-D2 $a$ (Å) | PBE $a$ (Å) |
| T-1 | 0.088 | 0.065 | 2.90 | 3.00 |
| T-2 | 0.088 | 0.065 | 2.90 | 3.00 |
| H-1 | 0.000 | 0.000 | 2.84 | 2.85 |
| H-2 | 0.000 | 0.000 | 2.84 | 2.85 |
| Mo₃C₂ (no termination) | | | | |
| | PBE-D2 $\Delta E$ (eV/atom) | PBE $\Delta E$ (eV/atom) | PBE-D2 $a$ (Å) | PBE $a$ (Å) |
| T-1 | 0.076 | 0.069 | 3.01 | 3.02 |
| T-2 | 0.076 | 0.069 | 3.01 | 3.02 |
| H1a-1 | 0.070 | 0.071 | 2.90 | 2.92 |
| H1a-2 | 0.070 | 0.072 | 2.90 | 2.92 |
| H1b-1 | 0.044 | 0.043 | 2.85 | 2.87 |
| H1b-2 | 0.044 | 0.043 | 2.85 | 2.87 |
| H2-1 | 0.000 | 0.000 | 2.85 | 2.87 |
| H2-2 | 0.000 | 0.000 | 2.85 | 2.87 |
| Mo₄C₃ (no termination) | | | | |
| | PBE-D2 $\Delta E$ (eV/atom) | PBE $\Delta E$ (eV/atom) | PBE-D2 $a$ (Å) | PBE $a$ (Å) |
| T-1 | 0.126 | 0.125 | 3.07 | 3.08 |
| T-2 | 0.126 | 0.125 | 3.08 | 3.08 |
| H1a-1 | 0.481 | 0.083 | 2.91 | 2.92 |
| H1a-2 | 0.074 | 0.083 | 2.91 | 2.92 |
| H1b-1 | 0.076 | 0.074 | 2.87 | 2.88 |
| H1b-2 | 0.076 | 0.074 | 2.87 | 2.88 |
| H2-1 | 0.000 | 0.000 | 2.86 | 2.88 |
| H2-2 | 0.000 | 0.000 | 2.86 | 2.88 |

**Table S7**. DFT-calculated energy difference ($\Delta E$) and DFT-optimized lattice constants ($a$) of all oxygen-terminated MoC-based 2D MXenes ($Mo_{n+1}C_nO_2$, where $n$ = 1-3) using PBE functional and PBE-D2 with vdW corrections. The energy difference is determined by: $\Delta E = E_{MX} - \min(E_{MX})$, where $E_{MX}$ is the ground state energy of each MXene structure and $\min(E_{MX})$ is the energy of the most stable MXene within the same group.

| | PBE-D2 $\Delta E$ (eV/atom) | PBE $\Delta E$ (eV/atom) | PBE-D2 $a$ (Å) | PBE $a$ (Å) |
|---|---|---|---|---|
| $Mo_2CO_2$ (oxygen termination) | | | | |
| T-1 | 0.331 | 0.279 | 2.95 | 3.09 |
| T-2 | 0.068 | 0.075 | 2.87 | 2.88 |
| H-1 | 0.194 | 0.194 | 2.88 | 2.90 |
| H-2 | 0.000 | 0.000 | 2.86 | 2.87 |
| $Mo_3C_2O_2$ (oxygen termination) | | | | |
| T-1 | 0.317 | 0.321 | 3.06 | 3.07 |
| T-2 | 0.165 | 0.166 | 2.90 | 2.92 |
| H1a-1 | 0.072 | 0.084 | 2.88 | 2.90 |
| H1a-2 | 0.269 | 0.278 | 2.95 | 2.96 |
| H1b-1 | 0.189 | 0.188 | 2.89 | 2.90 |
| H1b-2 | 0.066 | 0.065 | 2.87 | 2.89 |
| H2-1 | 0.000 | 0.000 | 2.87 | 2.89 |
| H2-2 | 0.165 | 0.163 | 2.86 | 2.91 |
| $Mo_4C_3O_2$ (oxygen termination) | | | | |
| T-1 | 0.304 | 0.314 | 3.04 | 3.05 |
| T-2 | 0.192 | 0.076 | 2.97 | 2.98 |
| H1a-1 | 0.075 | 0.090 | 2.90 | 2.91 |
| H1a-2 | 0.228 | 0.240 | 2.95 | 2.96 |
| H1b-1 | 0.177 | 0.176 | 2.89 | 2.91 |
| H1b-2 | 0.079 | 0.078 | 2.88 | 2.89 |
| H2-1 | 0.000 | 0.000 | 2.88 | 2.89 |
| H2-2 | 0.129 | 0.127 | 2.90 | 2.91 |

**Table S8**. DFT-calculated energy difference ($\Delta E$) and DFT-optimized lattice constants ($a$) of all fluoride-terminated MoC-based 2D MXenes ($Mo_{n+1}C_nF_2$, where $n$ = 1-3) using PBE functional and PBE-D2 with vdW corrections. The energy difference is determined by: $\Delta E = E_{MX} - \min(E_{MX})$, where $E_{MX}$ is the ground state energy of each MXene structure and $\min(E_{MX})$ is the energy of the most stable MXene within the same group.

| $Mo_2CF_2$ (fluoride termination) | | | | |
|---|---|---|---|---|
| | PBE-D2 $\Delta E$ (eV/atom) | PBE $\Delta E$ (eV/atom) | PBE-D2 $a$ (Å) | PBE $a$ (Å) |
| T-1 | 0.174 | 0.174 | 2.85 | 2.88 |
| T-2 | 0.097 | 0.081 | 2.96 | 2.98 |
| H-1 | 0.000 | 0.000 | 2.89 | 2.91 |
| H-2 | 0.056 | 0.054 | 2.92 | 2.93 |
| $Mo_3C_2F_2$ (fluoride termination) | | | | |
| | PBE-D2 $\Delta E$ (eV/atom) | PBE $\Delta E$ (eV/atom) | PBE-D2 $a$ (Å) | PBE $a$ (Å) |
| T-1 | 0.116 | 0.111 | 3.05 | 3.05 |
| T-2 | 0.113 | 0.106 | 3.00 | 3.01 |
| H1a-1 | 0.114 | 0.114 | 2.93 | 2.95 |
| H1a-2 | 0.129 | 0.137 | 2.87 | 2.89 |
| H1b-1 | 0.029 | 0.028 | 2.87 | 2.89 |
| H1b-2 | 0.027 | 0.024 | 2.90 | 2.91 |
| H2-1 | 0.043 | 0.040 | 2.88 | 2.90 |
| H2-2 | 0.000 | 0.000 | 2.86 | 2.88 |
| $Mo_4C_3F_2$ (fluoride termination) | | | | |
| | PBE-D2 $\Delta E$ (eV/atom) | PBE $\Delta E$ (eV/atom) | PBE-D2 $a$ (Å) | PBE $a$ (Å) |
| T-1 | 0.097 | 0.100 | 3.14 | 3.14 |
| T-2 | 0.129 | 0.127 | 3.04 | 3.05 |
| H1a-1 | 0.107 | 0.112 | 2.93 | 2.95 |
| H1a-2 | 0.120 | 0.131 | 2.89 | 2.91 |
| H1b-1 | 0.056 | 0.055 | 2.88 | 2.90 |
| H1b-2 | 0.055 | 0.052 | 2.89 | 2.91 |
| H2-1 | 0.027 | 0.024 | 2.89 | 2.91 |
| H2-2 | 0.000 | 0.000 | 2.87 | 2.89 |

**Table S9**. DFT-calculated energy difference ($\Delta E$) and DFT-optimized lattice constants ($a$) of all non-terminated TiN-based 2D MXenes ($Ti_{n+1}N_n$, where $n = 1$-3) using PBE functional and PBE-D2 with vdW corrections. The energy difference is determined by: $\Delta E = E_{MX} - \min(E_{MX})$, where $E_{MX}$ is the ground state energy of each MXene structure and $\min(E_{MX})$ is the energy of the most stable MXene within the same group.

| | PBE-D2 $\Delta E$ (eV/atom) | PBE $\Delta E$ (eV/atom) | PBE-D2 $a$ (Å) | PBE $a$ (Å) |
|---|---|---|---|---|
| Ti₂N (no termination) | | | | |
| T-1 | 0.000 | 0.000 | 2.99 | 2.98 |
| T-2 | 0.000 | 0.000 | 2.99 | 2.98 |
| H-1 | 0.318 | 0.316 | 2.89 | 2.89 |
| H-2 | 0.318 | 0.316 | 2.89 | 2.88 |
| Ti₃N₂ (no termination) | | | | |
| | PBE-D2 $\Delta E$ (eV/atom) | PBE $\Delta E$ (eV/atom) | PBE-D2 $a$ (Å) | PBE $a$ (Å) |
| T-1 | 0.000 | 0.000 | 3.04 | 3.04 |
| T-2 | 0.000 | 0.000 | 3.04 | 3.04 |
| H1a-1 | 0.045 | 0.042 | 3.02 | 3.02 |
| H1a-2 | 0.045 | 0.042 | 3.02 | 3.02 |
| H1b-1 | 0.312 | 0.303 | 2.91 | 2.91 |
| H1b-2 | 0.312 | 0.303 | 2.91 | 2.91 |
| H2-1 | 0.358 | 0.345 | 2.88 | 2.89 |
| H2-2 | 0.358 | 0.345 | 2.88 | 2.89 |
| Ti₄N₃ (no termination) | | | | |
| | PBE-D2 $\Delta E$ (eV/atom) | PBE $\Delta E$ (eV/atom) | PBE-D2 $a$ (Å) | PBE $a$ (Å) |
| T-1 | 0.000 | 0.000 | 2.99 | 2.99 |
| T-2 | 0.000 | 0.000 | 2.99 | 2.99 |
| H1a-1 | 0.097 | 0.092 | 2.97 | 2.97 |
| H1a-2 | 0.097 | 0.092 | 2.97 | 2.97 |
| H1b-1 | 0.323 | 0.309 | 2.93 | 2.94 |
| H1b-2 | 0.323 | 0.309 | 2.93 | 2.94 |
| H2-1 | 0.390 | 0.370 | 2.89 | 2.90 |
| H2-2 | 0.390 | 0.370 | 2.89 | 2.90 |

**Table S10**. DFT-calculated energy difference ($\Delta E$) and DFT-optimized lattice constants ($a$) of all oxygen-terminated TiN-based 2D MXenes ($Ti_{n+1}N_nO_2$, where $n$ = 1-3) using PBE functional and PBE-D2 with vdW corrections. The energy difference is determined by: $\Delta E = E_{MX} - \min(E_{MX})$, where $E_{MX}$ is the ground state energy of each MXene structure and $\min(E_{MX})$ is the energy of the most stable MXene within the same group.

| | PBE-D2 $\Delta E$ (eV/atom) | PBE $\Delta E$ (eV/atom) | PBE-D2 $a$ (Å) | PBE $a$ (Å) |
|---|---|---|---|---|
| Ti_2NO_2 (oxygen termination) | | | | |
| T-1 | 0.000 | 0.000 | 3.00 | 3.00 |
| T-2 | 0.254 | 0.248 | 2.91 | 2.92 |
| H-1 | 0.079 | 0.069 | 3.01 | 3.01 |
| H-2 | 0.335 | 0.315 | 2.92 | 2.92 |
| Ti_3N_2O_2 (oxygen termination) | | | | |
| | PBE-D2 $\Delta E$ (eV/atom) | PBE $\Delta E$ (eV/atom) | PBE-D2 $a$ (Å) | PBE $a$ (Å) |
| T-1 | 0.000 | 0.000 | 3.01 | 3.01 |
| T-2 | 0.184 | 0.181 | 2.95 | 2.95 |
| H1a-1 | 0.212 | 0.207 | 2.92 | 2.93 |
| H1a-2 | 0.015 | 0.012 | 2.98 | 2.98 |
| H1b-1 | 0.106 | 0.094 | 3.02 | 3.02 |
| H1b-2 | 0.323 | 0.303 | 2.94 | 2.94 |
| H2-1 | 0.359 | 0.336 | 2.91 | 2.92 |
| H2-2 | 0.130 | 0.113 | 2.98 | 2.99 |
| Ti_4N_3O_2 (oxygen termination) | | | | |
| | PBE-D2 $\Delta E$ (eV/atom) | PBE $\Delta E$ (eV/atom) | PBE-D2 $a$ (Å) | PBE $a$ (Å) |
| T-1 | 0.000 | 0.000 | 3.00 | 3.00 |
| T-2 | 0.131 | 0.129 | 2.96 | 2.97 |
| H1a-1 | 0.185 | 0.179 | 2.93 | 2.93 |
| H1a-2 | 0.045 | 0.041 | 2.96 | 2.97 |
| H1b-1 | 0.141 | 0.125 | 2.99 | 3.00 |
| H1b-2 | 0.306 | 0.286 | 2.95 | 2.95 |
| H2-1 | 0.361 | 0.336 | 2.91 | 2.92 |
| H2-2 | 0.186 | 0.165 | 2.95 | 2.96 |

**Table S11**. DFT-calculated energy difference ($\Delta E$) and DFT-optimized lattice constants ($a$) of all fluoride-terminated TiN-based 2D MXenes ($Ti_{n+1}N_nF_2$, where $n$ = 1-3) using PBE functional and PBE-D2 with vdW corrections. The energy difference is determined by: $\Delta E = E_{MX} - \min(E_{MX})$, where $E_{MX}$ is the ground state energy of each MXene structure and $\min(E_{MX})$ is the energy of the most stable MXene within the same group.

| $Ti_2NF_2$ (fluoride termination) | | | | |
|---|---|---|---|---|
| | PBE-D2 $\Delta E$ (eV/atom) | PBE $\Delta E$ (eV/atom) | PBE-D2 $a$ (Å) | PBE $a$ (Å) |
| T-1 | 0.000 | 0.000 | 3.06 | 3.06 |
| T-2 | 0.084 | 0.085 | 2.90 | 2.91 |
| H-1 | 0.207 | 0.204 | 2.90 | 2.91 |
| H-2 | 0.164 | 0.159 | 2.88 | 2.89 |
| $Ti_3N_2F_2$ (fluoride termination) | | | | |
| | PBE-D2 $\Delta E$ (eV/atom) | PBE $\Delta E$ (eV/atom) | PBE-D2 $a$ (Å) | PBE $a$ (Å) |
| T-1 | 0.011 | 0.012 | 3.02 | 3.03 |
| T-2 | 0.000 | 0.000 | 2.94 | 2.94 |
| H1a-1 | 0.040 | 0.038 | 2.92 | 2.93 |
| H1a-2 | 0.064 | 0.064 | 2.98 | 2.99 |
| H1b-1 | 0.188 | 0.180 | 2.93 | 2.94 |
| H1b-2 | 0.161 | 0.151 | 2.91 | 2.92 |
| H2-1 | 0.186 | 0.173 | 2.89 | 2.90 |
| H2-2 | 0.214 | 0.203 | 2.91 | 2.92 |
| $Ti_4N_3F_2$ (fluoride termination) | | | | |
| | PBE-D2 $\Delta E$ (eV/atom) | PBE $\Delta E$ (eV/atom) | PBE-D2 $a$ (Å) | PBE $a$ (Å) |
| T-1 | 0.000 | 0.000 | 3.02 | 3.02 |
| T-2 | 0.020 | 0.019 | 2.95 | 2.96 |
| H1a-1 | 0.073 | 0.069 | 2.92 | 2.93 |
| H1a-2 | 0.072 | 0.068 | 2.99 | 3.00 |
| H1b-1 | 0.211 | 0.200 | 2.95 | 2.95 |
| H1b-2 | 0.194 | 0.181 | 2.92 | 2.93 |
| H2-1 | 0.241 | 0.223 | 2.89 | 2.90 |
| H2-2 | 0.261 | 0.245 | 2.91 | 2.92 |

**Table S12**. DFT-calculated energy difference ($\Delta E$) and DFT-optimized lattice constants ($a$) of all non-terminated MoN-based 2D MXenes (Mo$_{n+1}$N$_n$, where $n$ = 1-3) using PBE functional and PBE-D2 with vdW corrections. The energy difference is determined by: $\Delta E = E_{MX} - \min(E_{MX})$, where $E_{MX}$ is the ground state energy of each MXene structure and $\min(E_{MX})$ is the energy of the most stable MXene within the same group.

| Mo$_2$N (no termination) | | | | |
|---|---|---|---|---|
| | PBE-D2 $\Delta E$ (eV/atom) | PBE $\Delta E$ (eV/atom) | PBE-D2 $a$ (Å) | PBE $a$ (Å) |
| T-1 | 0.088 | 0.095 | 2.90 | 2.82 |
| T-2 | 0.088 | 0.095 | 2.90 | 2.82 |
| H-1 | 0.000 | 0.000 | 2.84 | 2.83 |
| H-2 | 0.000 | 0.000 | 2.84 | 2.83 |
| Mo$_3$N$_2$ (no termination) | | | | |
| | PBE-D2 $\Delta E$ (eV/atom) | PBE $\Delta E$ (eV/atom) | PBE-D2 $a$ (Å) | PBE $a$ (Å) |
| T-1 | 0.375 | 0.358 | 3.08 | 3.08 |
| T-2 | 0.190 | 0.206 | 2.78 | 2.79 |
| H1a-1 | 0.178 | 0.189 | 2.82 | 2.84 |
| H1a-2 | 0.178 | 0.189 | 2.82 | 2.84 |
| H1b-1 | 0.000 | 0.000 | 2.81 | 2.83 |
| H1b-2 | 0.000 | 0.000 | 2.82 | 2.83 |
| H2-1 | 0.056 | 0.054 | 2.82 | 2.83 |
| H2-2 | 0.056 | 0.054 | 2.82 | 2.83 |
| Mo$_4$N$_3$ (no termination) | | | | |
| | PBE-D2 $\Delta E$ (eV/atom) | PBE $\Delta E$ (eV/atom) | PBE-D2 $a$ (Å) | PBE $a$ (Å) |
| T-1 | 0.272 | 0.000 | 3.23 | 2.99 |
| T-2 | 0.203 | 0.307 | 2.78 | 2.80 |
| H1a-1 | 0.203 | 0.296 | 2.85 | 2.87 |
| H1a-2 | 0.203 | 0.296 | 2.85 | 2.87 |
| H1b-1 | 0.000 | 0.084 | 2.82 | 2.83 |
| H1b-2 | 0.000 | 0.084 | 2.82 | 2.83 |
| H2-1 | 0.064 | 0.146 | 2.83 | 2.84 |
| H2-2 | 0.064 | 0.146 | 2.83 | 2.84 |

**Table S13**. DFT-calculated energy difference ($\Delta E$) and DFT-optimized lattice constants ($a$) of all oxygen-terminated MoN-based 2D MXenes ($Mo_{n+1}N_nO_2$, where $n$ = 1-3) using PBE functional and PBE-D2 with vdW corrections. The energy difference is determined by: $\Delta E = E_{MX} - \min(E_{MX})$, where $E_{MX}$ is the ground state energy of each MXene structure and $\min(E_{MX})$ is the energy of the most stable MXene within the same group.

| Mo₂NO₂ (oxygen termination) | | | | |
|---|---|---|---|---|
| | PBE-D2 $\Delta E$ (eV/atom) | PBE $\Delta E$ (eV/atom) | PBE-D2 $a$ (Å) | PBE $a$ (Å) |
| T-1 | 0.259 | 0.272 | 2.85 | 2.86 |
| T-2 | 0.259 | 0.061 | 2.85 | 2.87 |
| H-1 | 0.096 | 0.099 | 2.87 | 2.88 |
| H-2 | 0.000 | 0.000 | 2.87 | 2.88 |
| Mo₃N₂O₂ (oxygen termination) | | | | |
| | PBE-D2 $\Delta E$ (eV/atom) | PBE $\Delta E$ (eV/atom) | PBE-D2 $a$ (Å) | PBE $a$ (Å) |
| T-1 | 0.328 | 0.329 | 3.14 | 3.14 |
| T-2 | 0.104 | 0.119 | 2.83 | 2.84 |
| H1a-1 | 0.142 | 0.153 | 2.87 | 2.88 |
| H1a-2 | 0.241 | 0.258 | 2.82 | 2.84 |
| H1b-1 | 0.102 | 0.104 | 2.85 | 2.87 |
| H1b-2 | 0.000 | 0.000 | 2.83 | 2.85 |
| H2-1 | 0.072 | 0.071 | 2.87 | 2.88 |
| H2-2 | 0.101 | 0.101 | 2.84 | 2.86 |
| Mo₄N₃O₂ (oxygen termination) | | | | |
| | PBE-D2 $\Delta E$ (eV/atom) | PBE $\Delta E$ (eV/atom) | PBE-D2 $a$ (Å) | PBE $a$ (Å) |
| T-1 | 0.263 | 0.270 | 3.19 | 3.19 |
| T-2 | 0.133 | 0.150 | 2.82 | 2.83 |
| H1a-1 | 0.170 | 0.182 | 2.87 | 2.89 |
| H1a-2 | 0.238 | 0.254 | 2.84 | 2.85 |
| H1b-1 | 0.079 | 0.082 | 2.85 | 2.87 |
| H1b-2 | 0.000 | 0.000 | 2.83 | 2.85 |
| H2-1 | 0.078 | 0.076 | 2.85 | 2.87 |
| H2-2 | 0.086 | 0.085 | 2.85 | 2.87 |

**Table S14**. DFT-calculated energy difference ($\Delta E$) and DFT-optimized lattice constants ($a$) of all fluoride-terminated MoN-based 2D MXenes ($Mo_{n+1}N_nF_2$, where $n$ = 1-3) using PBE functional and PBE-D2 with vdW corrections. The energy difference is determined by: $\Delta E = E_{MX} - \min(E_{MX})$, where $E_{MX}$ is the ground state energy of each MXene structure and $\min(E_{MX})$ is the energy of the most stable MXene within the same group.

| Mo₂NF₂ (fluoride termination) | | | | |
|---|---|---|---|---|
| | PBE-D2 $\Delta E$ (eV/atom) | PBE $\Delta E$ (eV/atom) | PBE-D2 $a$ (Å) | PBE $a$ (Å) |
| T-1 | 0.174 | 0.114 | 2.85 | 2.78 |
| T-2 | 0.097 | 0.228 | 2.96 | 3.00 |
| H-1 | 0.000 | 0.000 | 2.89 | 2.80 |
| H-2 | 0.056 | 0.151 | 2.92 | 2.98 |
| Mo₃N₂F₂ (fluoride termination) | | | | |
| | PBE-D2 $\Delta E$ (eV/atom) | PBE $\Delta E$ (eV/atom) | PBE-D2 $a$ (Å) | PBE $a$ (Å) |
| T-1 | 0.245 | 0.237 | 3.21 | 3.20 |
| T-2 | 0.334 | 0.316 | 2.99 | 3.02 |
| H1a-1 | 0.310 | 0.305 | 2.88 | 2.91 |
| H1a-2 | 0.166 | 0.173 | 2.82 | 2.84 |
| H1b-1 | 0.000 | 0.000 | 2.82 | 2.84 |
| H1b-2 | 0.128 | 0.125 | 2.83 | 2.85 |
| H2-1 | 0.165 | 0.160 | 2.83 | 2.84 |
| H2-2 | 0.021 | 0.019 | 2.82 | 2.84 |
| Mo₄N₃F₂ (fluoride termination) | | | | |
| | PBE-D2 $\Delta E$ (eV/atom) | PBE $\Delta E$ (eV/atom) | PBE-D2 $a$ (Å) | PBE $a$ (Å) |
| T-1 | 0.132 | 0.124 | 3.22 | 3.23 |
| T-2 | 0.271 | 0.284 | 2.79 | 2.81 |
| H1a-1 | 0.290 | 0.285 | 2.90 | 2.93 |
| H1a-2 | 0.194 | 0.201 | 2.84 | 2.87 |
| H1b-1 | 0.000 | 0.000 | 2.82 | 2.84 |
| H1b-2 | 0.089 | 0.087 | 2.83 | 2.85 |
| H2-1 | 0.150 | 0.145 | 2.84 | 2.86 |
| H2-2 | 0.046 | 0.043 | 2.83 | 2.83 |

**Table S15**. DFT-calculated energy difference ($\Delta E$) and DFT-optimized lattice constants ($a$) of all fluoride-terminated HfC- and HfN-based 2D MXenes using PBE-D2 functional with vdW corrections. The energy difference is determined by: $\Delta E = E_{MX} - \min(E_{MX})$, where $E_{MX}$ is the ground state energy of each MXene structure and $\min(E_{MX})$ is the energy of the most stable MXene within the same group.

| | Hf$_2$CF$_2$ | | Hf$_2$NF$_2$ | |
|---|---|---|---|---|
| | $\Delta E$ (eV/atom) | $a$ (Å) | $\Delta E$ (eV/atom) | $a$ (Å) |
| T-1 | 0.000 | 3.26 | 0.028 | 3.25 |
| T-2 | 0.064 | 3.15 | 0.000 | 3.06 |
| H-1 | 0.259 | 3.24 | 0.160 | 2.98 |
| H-2 | 0.305 | 3.11 | 0.051 | 3.02 |
| | Hf$_3$C$_2$F$_2$ | | Hf$_3$N$_2$F$_2$ | |
| | $\Delta E$ (eV/atom) | $a$ (Å) | $\Delta E$ (eV/atom) | $a$ (Å) |
| T-1 | 0.000 | 3.29 | 0.099 | 3.20 |
| T-2 | 0.088 | 3.21 | 0.000 | 3.10 |
| H1a-1 | 0.114 | 3.20 | 0.026 | 3.08 |
| H1a-2 | 0.003 | 3.29 | 0.131 | 3.09 |
| H1b-1 | 0.310 | 3.25 | 0.247 | 3.01 |
| H1b-2 | 0.366 | 3.16 | 0.164 | 3.04 |
| H2-1 | 0.419 | 3.14 | 0.162 | 3.02 |
| H2-2 | 0.337 | 3.24 | 0.258 | 3.01 |
| | Hf$_4$C$_3$F$_2$ | | Hf$_4$N$_3$F$_2$ | |
| | $\Delta E$ (eV/atom) | $a$ (Å) | $\Delta E$ (eV/atom) | $a$ (Å) |
| T-1 | 0.000 | 3.30 | 0.041 | 3.21 |
| T-2 | 0.070 | 3.23 | 0.000 | 3.11 |
| H1a-1 | 0.121 | 3.22 | 0.019 | 3.08 |
| H1a-2 | 0.025 | 3.28 | 0.088 | 3.12 |
| H1b-1 | 0.337 | 3.24 | 0.222 | 3.03 |
| H1b-2 | 0.396 | 3.18 | 0.159 | 3.05 |
| H2-1 | 0.472 | 3.15 | 0.169 | 3.02 |
| H2-2 | 0.396 | 3.23 | 0.241 | 3.00 |

**Table S16**. DFT-calculated energy difference ($\Delta E$) and DFT-optimized lattice constants ($a$) of all fluoride-terminated ReC- and ReN-based 2D MXenes using PBE-D2 functional with vdW corrections. The energy difference is determined by: $\Delta E = E_{MX} - \min(E_{MX})$, where $E_{MX}$ is the ground state energy of each MXene structure and $\min(E_{MX})$ is the energy of the most stable MXene within the same group.

| | Re$_2$CF$_2$ | | Re$_2$NF$_2$ | |
|---|---|---|---|---|
| | $\Delta E$ (eV/atom) | $a$ (Å) | $\Delta E$ (eV/atom) | $a$ (Å) |
| T-1 | 0.329 | 3.34 | 0.310 | 3.39 |
| T-2 | 0.359 | 3.03 | 0.306 | 2.65 |
| H-1 | 0.000 | 2.77 | 0.000 | 2.73 |
| H-2 | 0.135 | 2.79 | 0.266 | 3.07 |
| | Re$_3$C$_2$F$_2$ | | Re$_3$N$_2$F$_2$ | |
| | $\Delta E$ (eV/atom) | $a$ (Å) | $\Delta E$ (eV/atom) | $a$ (Å) |
| T-1 | 0.845 | 3.19 | 0.977 | 3.31 |
| T-2 | 0.913 | 3.06 | 1.216 | 3.18 |
| H1a-1 | 0.889 | 2.85 | 0.968 | 2.67 |
| H1a-2 | 0.000 | 3.29 | 0.823 | 3.39 |
| H1b-1 | 0.399 | 2.79 | 0.000 | 2.73 |
| H1b-2 | 0.493 | 2.80 | 0.774 | 2.72 |
| H2-1 | 0.495 | 2.80 | 0.827 | 2.71 |
| H2-2 | 0.401 | 2.80 | 0.699 | 2.71 |
| | Re$_4$C$_3$F$_2$ | | Re$_4$N$_3$F$_2$ | |
| | $\Delta E$ (eV/atom) | $a$ (Å) | $\Delta E$ (eV/atom) | $a$ (Å) |
| T-1 | 0.431 | 3.24 | 0.444 | 3.29 |
| T-2 | 0.565 | 3.17 | 0.330 | 2.69 |
| H1a-1 | 0.484 | 2.86 | 0.665 | 3.08 |
| H1a-2 | 0.405 | 2.82 | 0.271 | 2.68 |
| H1b-1 | 0.015 | 2.81 | 0.000 | 2.73 |
| H1b-2 | 0.082 | 2.81 | 0.123 | 2.73 |
| H2-1 | 0.082 | 2.81 | 0.230 | 2.73 |
| H2-2 | 0.000 | 2.81 | 0.125 | 2.73 |

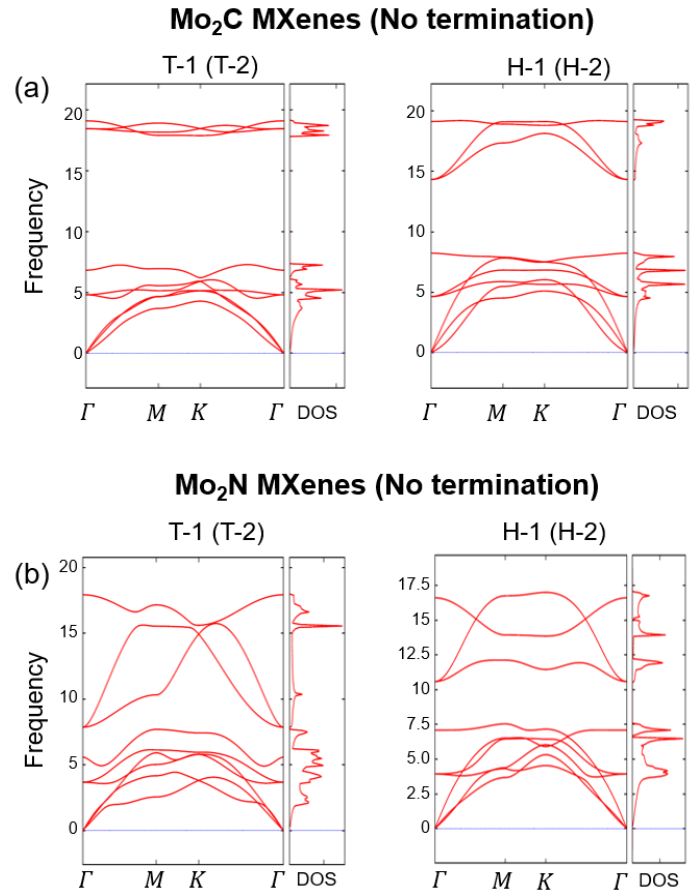

Fig. S1. Phonon spectra and vibrational density of states (vDOS) of (a) four Mo$_2$C MXenes and (b) four Mo$_2$N MXenes. Without termination atoms, two bulk derived MXenes are identical.

**Mo₃C₂ MXenes (No termination)**

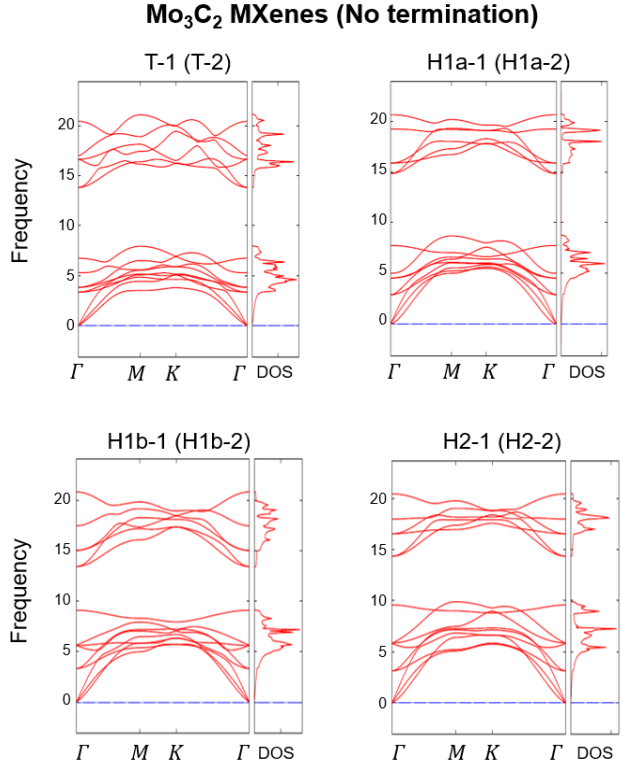

Fig. S2. Phonon spectra and vibrational density of states (vDOS) of eight types of Mo₃C₂ MXenes. Without termination elements, two bulk derived MXenes are identical.

**Mo₃N₂ MXenes (No termination)**

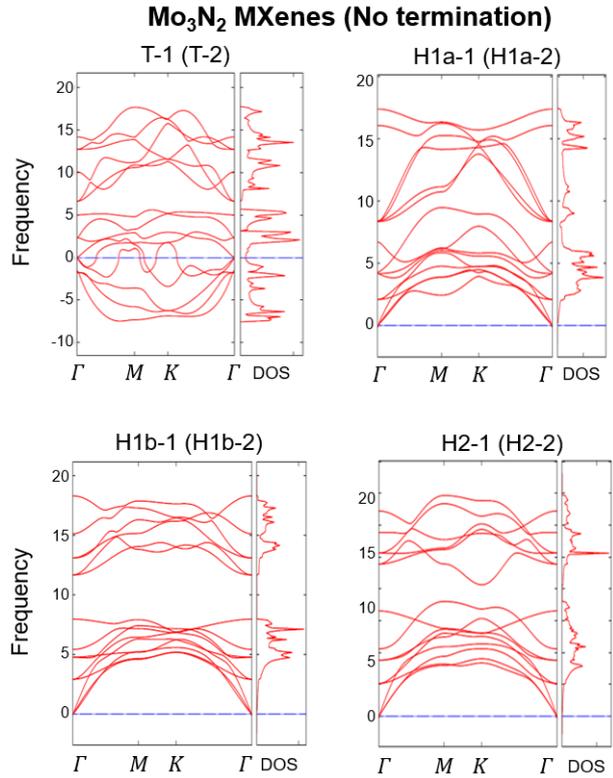

Fig. S3. Phonon spectra and vibrational density of states (vDOS) of eight types of Mo₃N₂ MXenes. Without termination elements, two bulk derived MXenes are identical.

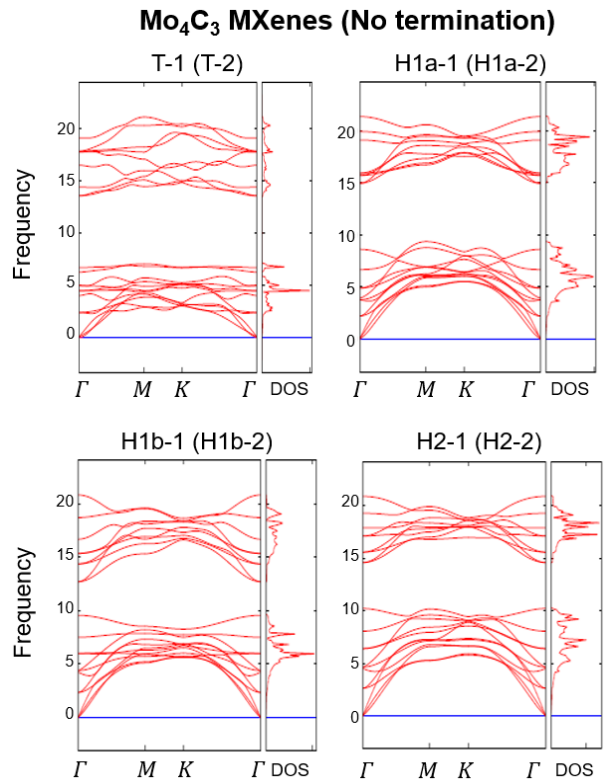

Fig. S4. Phonon spectra and vibrational density of states (vDOS) of eight types of Mo₄C₃ MXenes. Without termination elements, two bulk derived MXenes are identical.

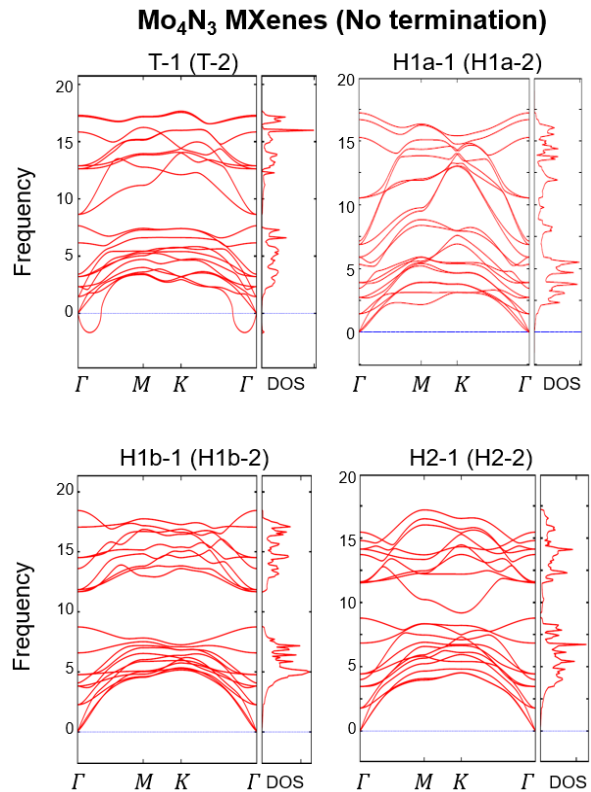

Fig. S5. Phonon spectra and vibrational density of states (vDOS) of eight types of Mo₄N₃ MXenes. Without termination elements, two bulk derived MXenes are identical.

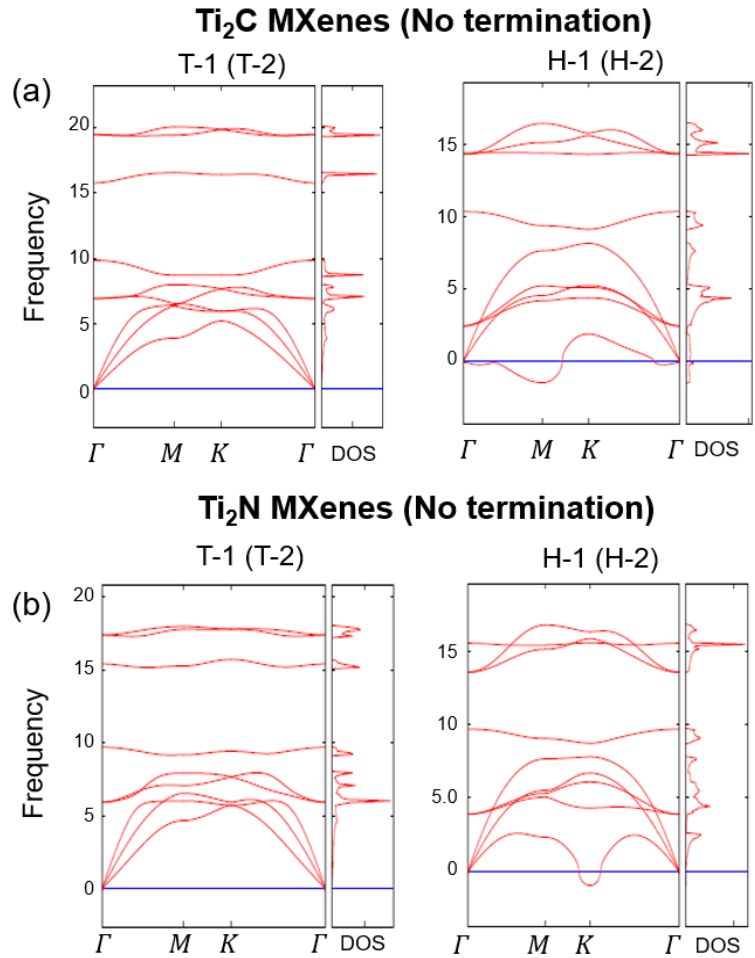

Fig. S6. Phonon spectra and vibrational density of states (vDOS) of (a) four Ti₂C MXenes and (b) four Ti₂N MXenes. Without termination atoms, two bulk derived MXenes are identical.

**Ti₃C₂ MXenes (No termination)**

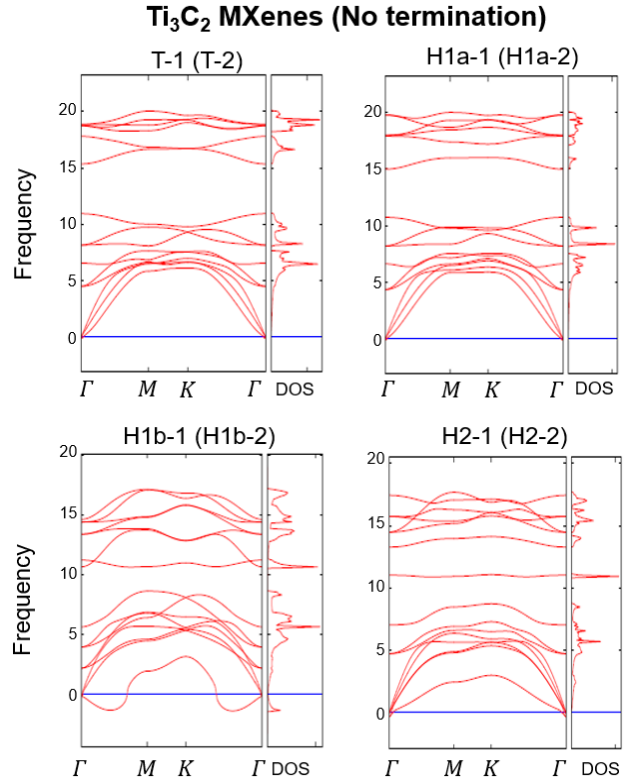

Fig. S7. Phonon spectra and vibrational density of states (vDOS) of eight types of Ti₃C₂ MXenes. Without termination atoms, two bulk derived MXenes are identical.

**Ti₃N₂ MXenes (No termination)**

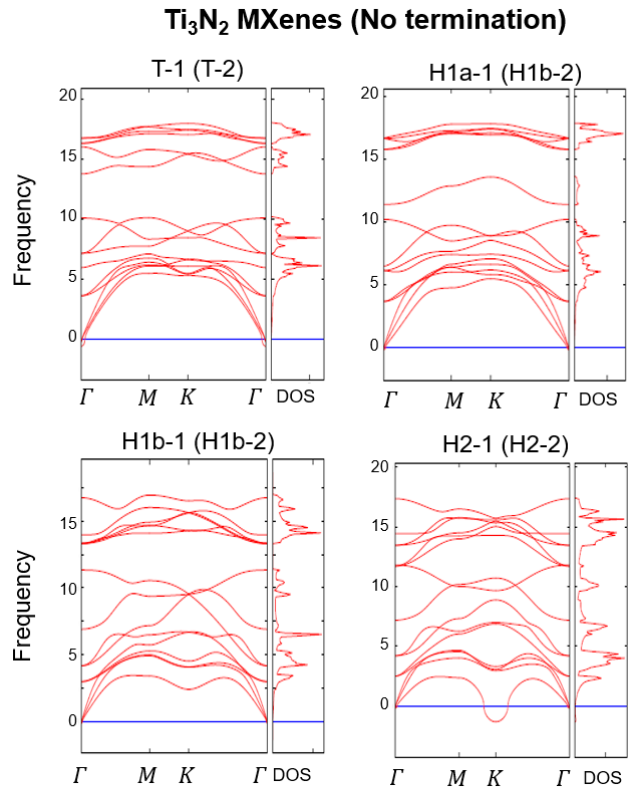

Fig. S8. Phonon spectra and vibrational density of states (DOSs) of eight types of Ti₃N₂ MXenes. Without termination atoms, two bulk derived MXenes are identical.

**Ti₄C₃ MXenes (No termination)**

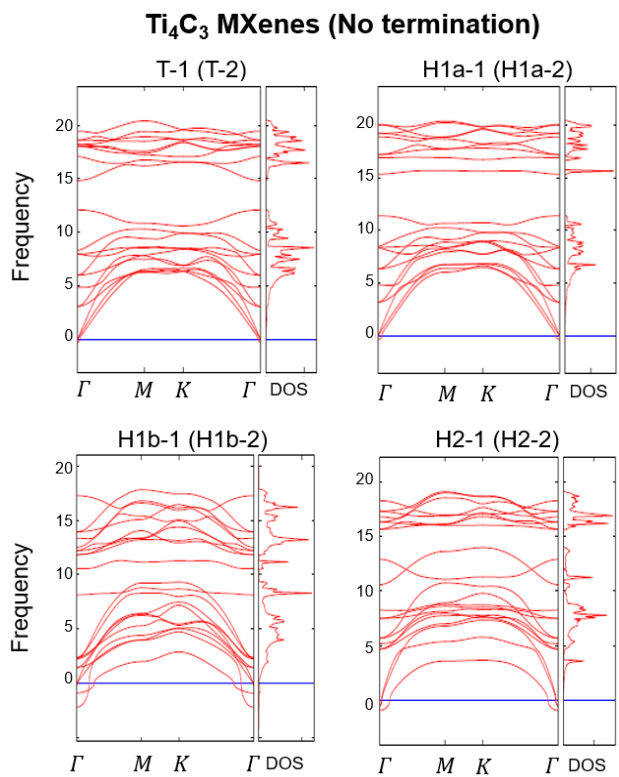

Fig. S9. Phonon spectra and vibrational density of states (vDOS) of eight types of Ti₄C₃ MXenes. Without termination atoms, two bulk derived MXenes are identical.

**Ti₄N₃ MXenes (No termination)**

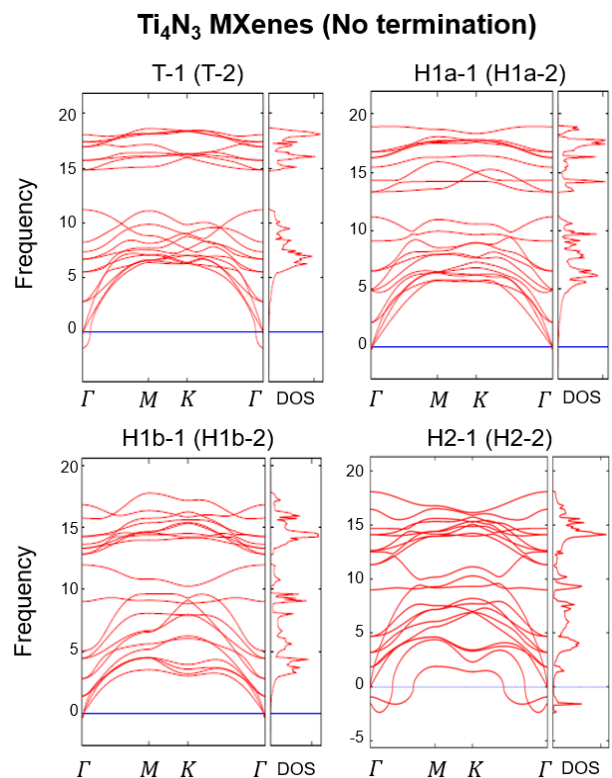

Fig. S10. Phonon spectra and vibrational density of states (DOSs) of eight types of Ti₄N₃ MXenes. Without termination atoms, two bulk derived MXenes are identical.

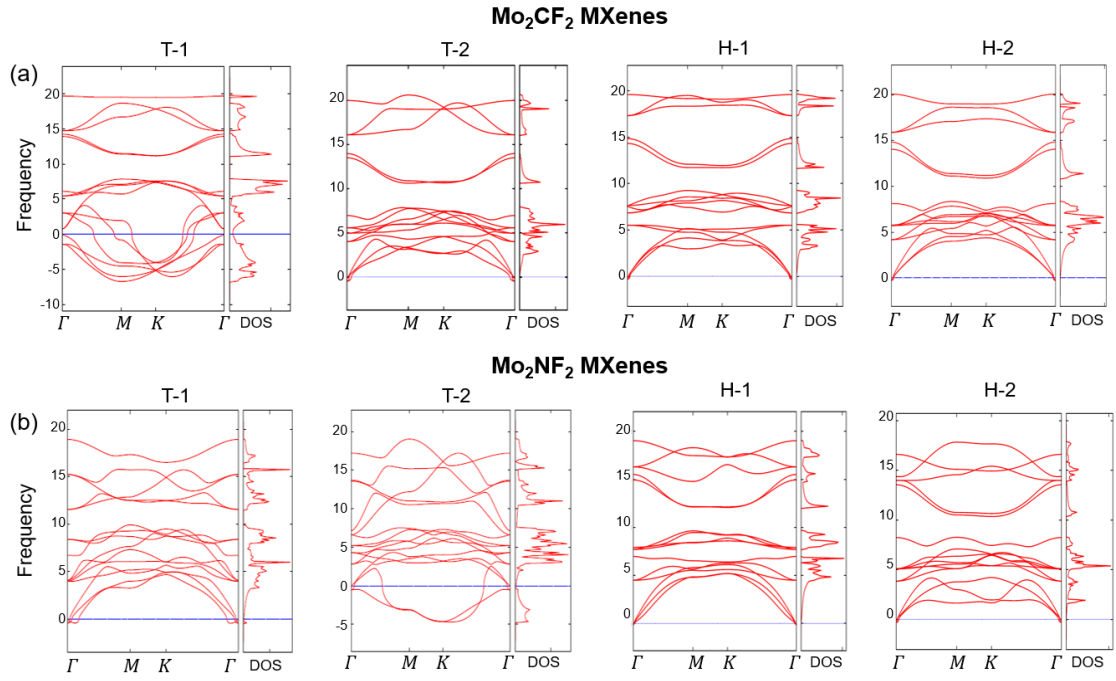

Fig. S11. Phonon spectra and vibrational density of states (vDOS) of (a) four $Mo_2CF_2$ MXenes and (b) four $Mo_2NF_2$ MXenes, both terminated by fluorine atoms.

**Mo₃C₂F₂ MXenes**

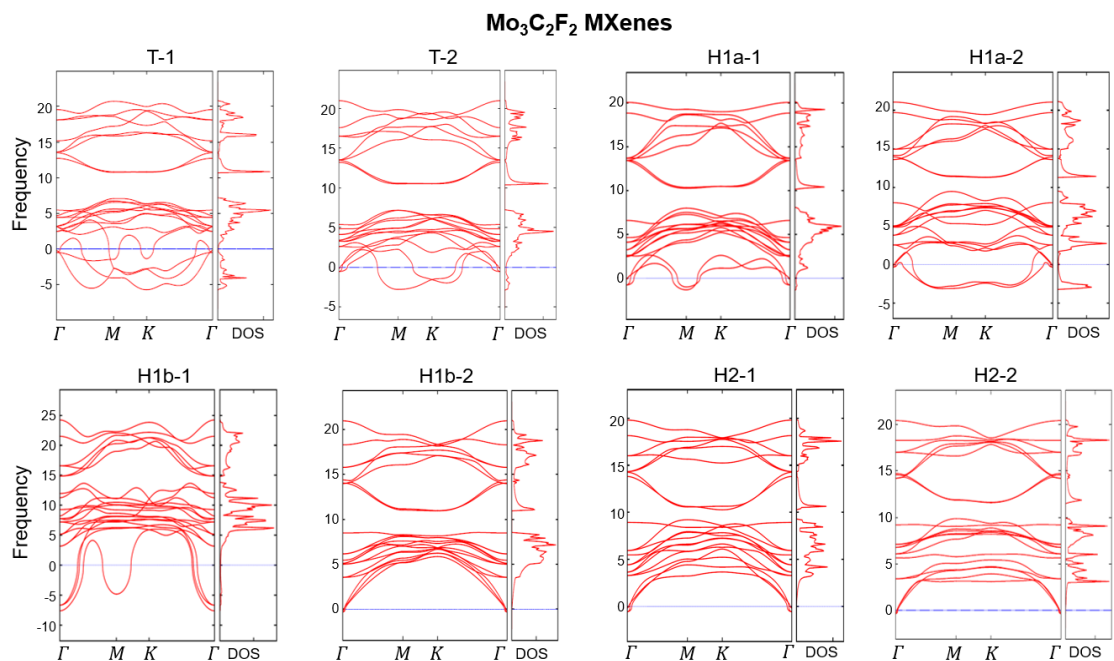

Fig. S12. Phonon spectra and vibrational density of states (vDOS) of eight types of Mo₃C₂F₂-based MXenes terminated by fluorine atoms.

**Mo₃N₂F₂ MXenes**

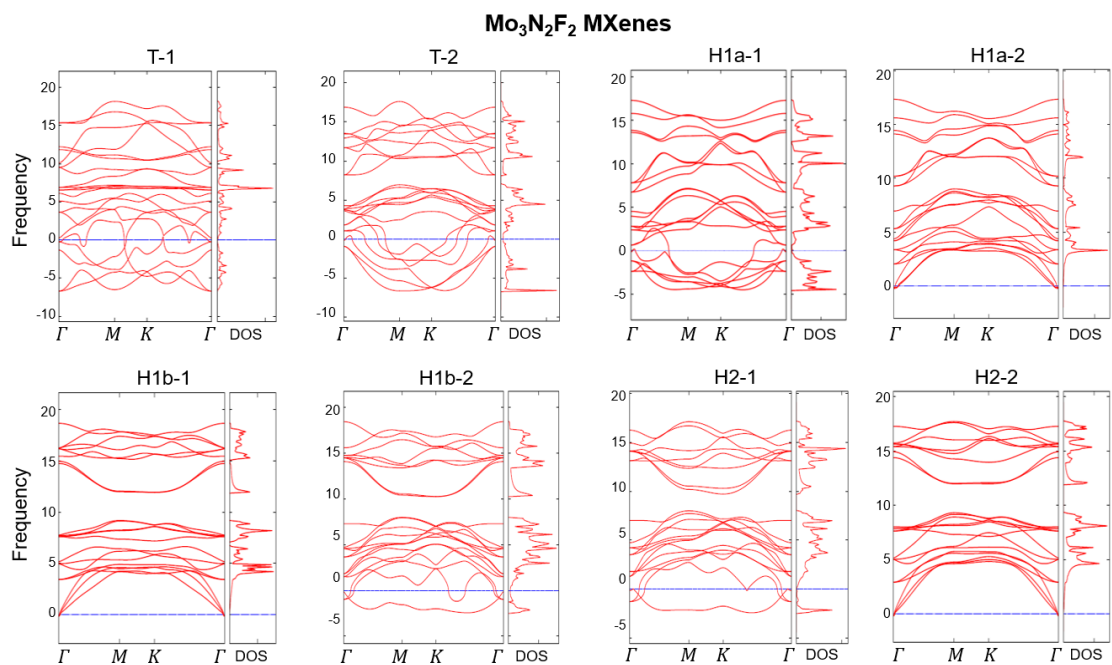

Fig. S13. Phonon spectra and vibrational density of states (vDOS) of eight types of Mo₃N₂F₂ MXenes terminated by fluorine atoms.

**Mo₄C₃F₂ MXenes**

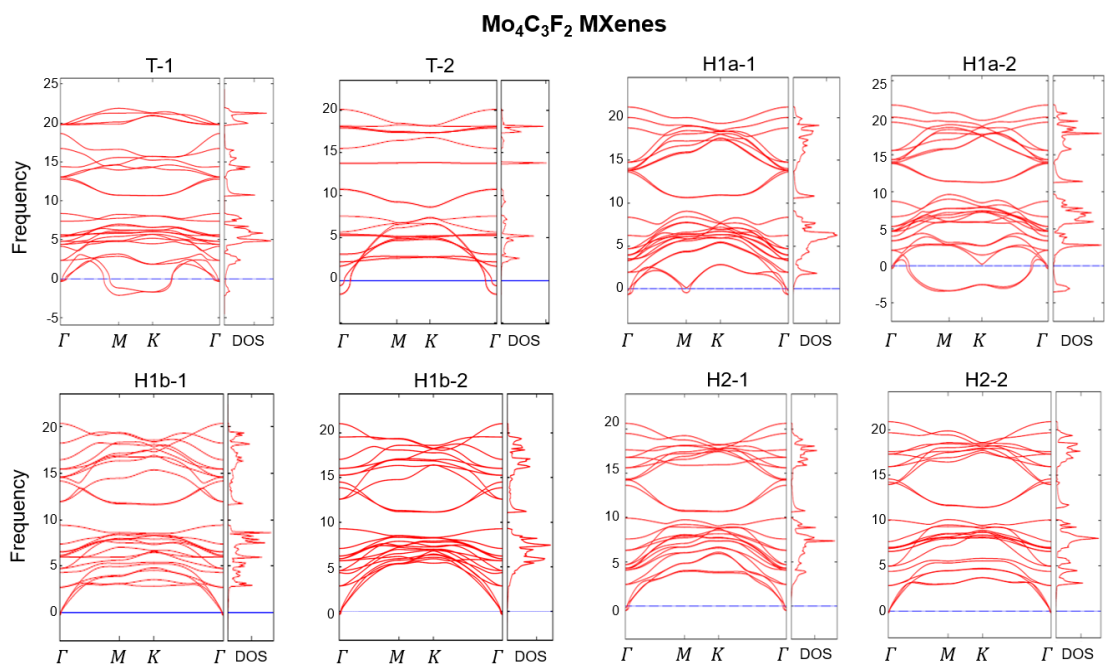

Fig. S14. Phonon spectra and vibrational density of states (vDOS) of eight types of Mo₄C₃F₂ MXenes terminated by fluorine atoms.

**Mo₄N₃F₂ MXenes**

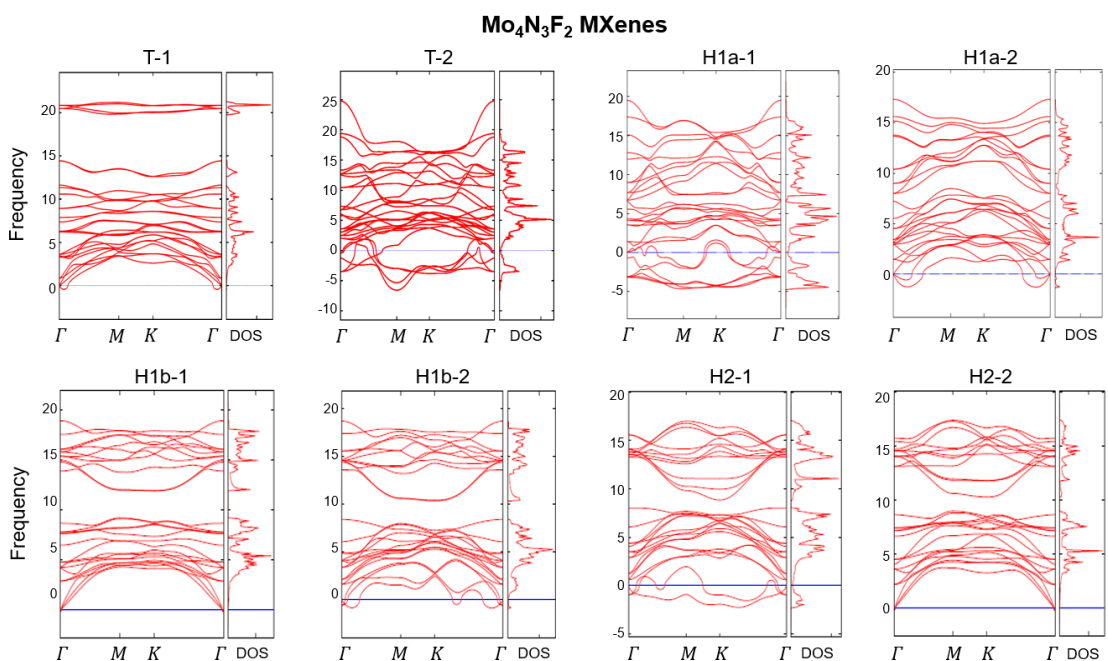

Fig. S15. Phonon spectra and vibrational density of states (vDOS) of eight types of Mo₄N₃F₂ MXenes terminated by fluorine atoms.

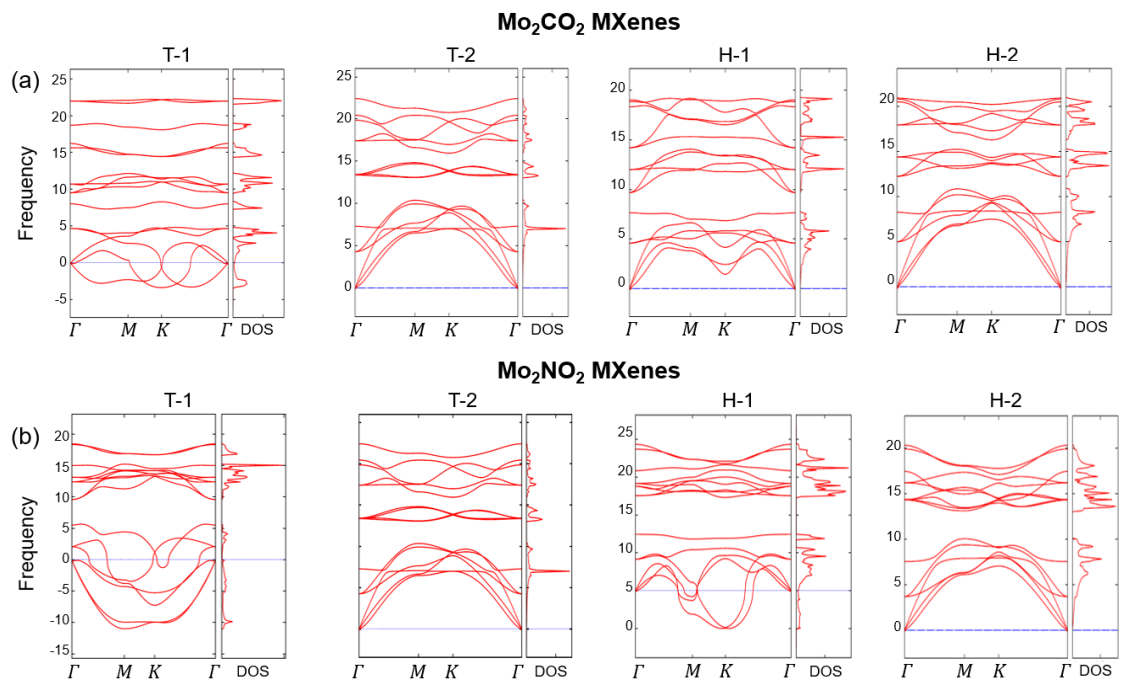

Fig. S16. Phonon spectra and vibrational density of states (vDOS) of (a) four Mo$_2$CO$_2$ MXenes and (b) four Mo$_2$NO$_2$ MXenes, both terminated by oxygen atoms.

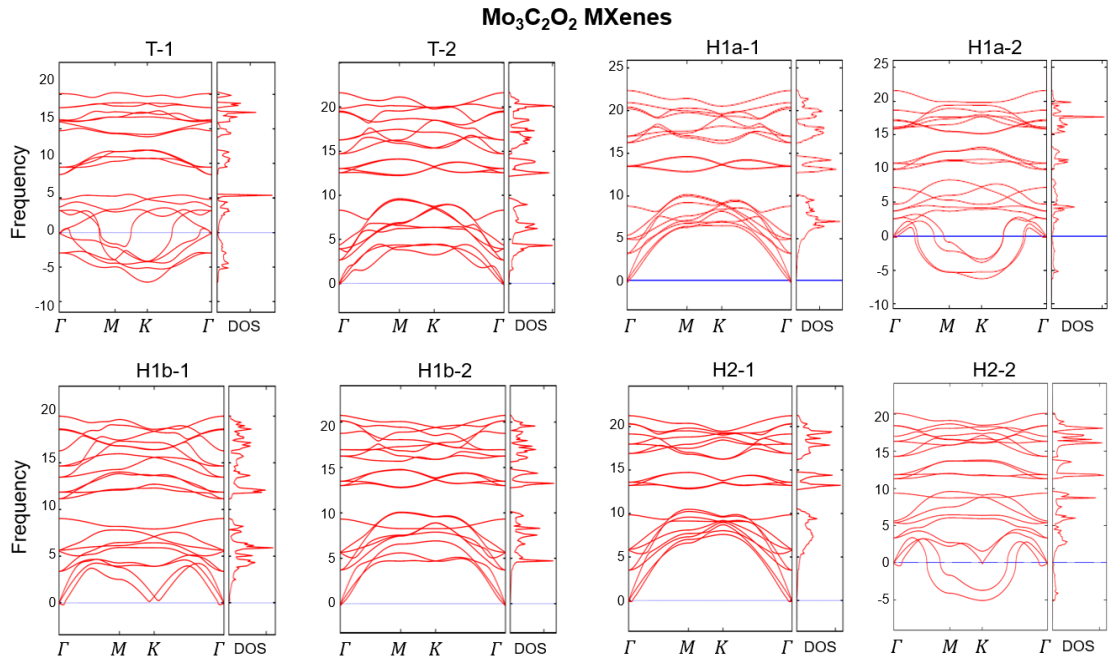

Fig. S17. Phonon spectra and vibrational density of states (vDOS) of eight types of Mo₃C₂O₂-based MXenes terminated by oxygen atoms.

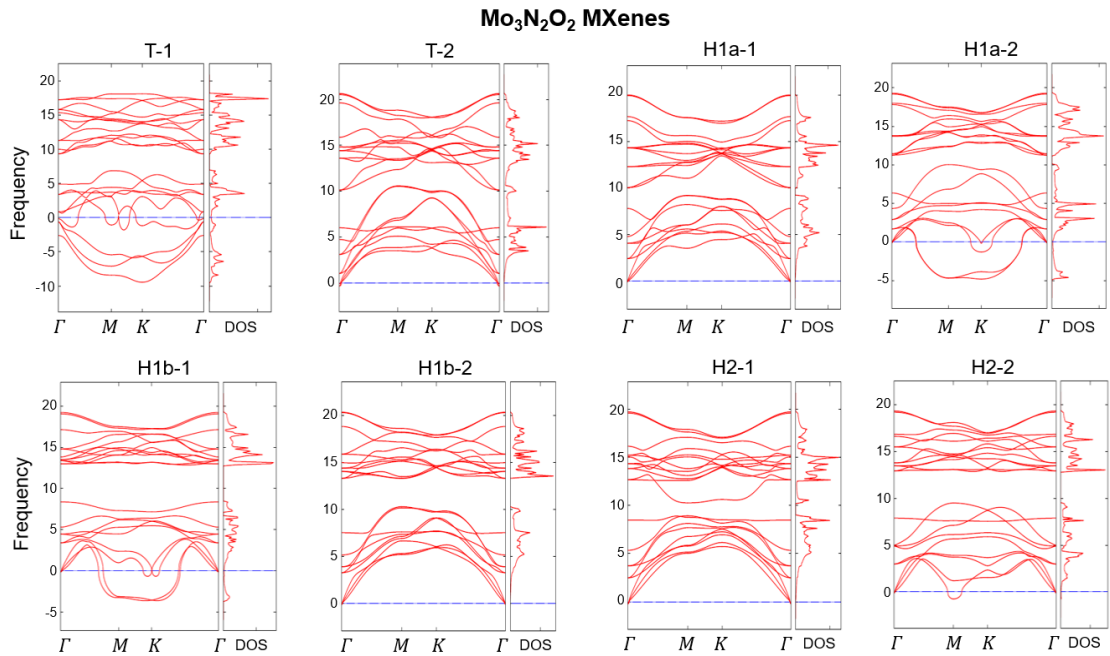

Fig. S18. Phonon spectra and vibrational density of states (vDOS) of eight types of Mo₃N₂O₂-based MXenes terminated by oxygen atoms.

**Mo₄C₃O₂ MXenes**

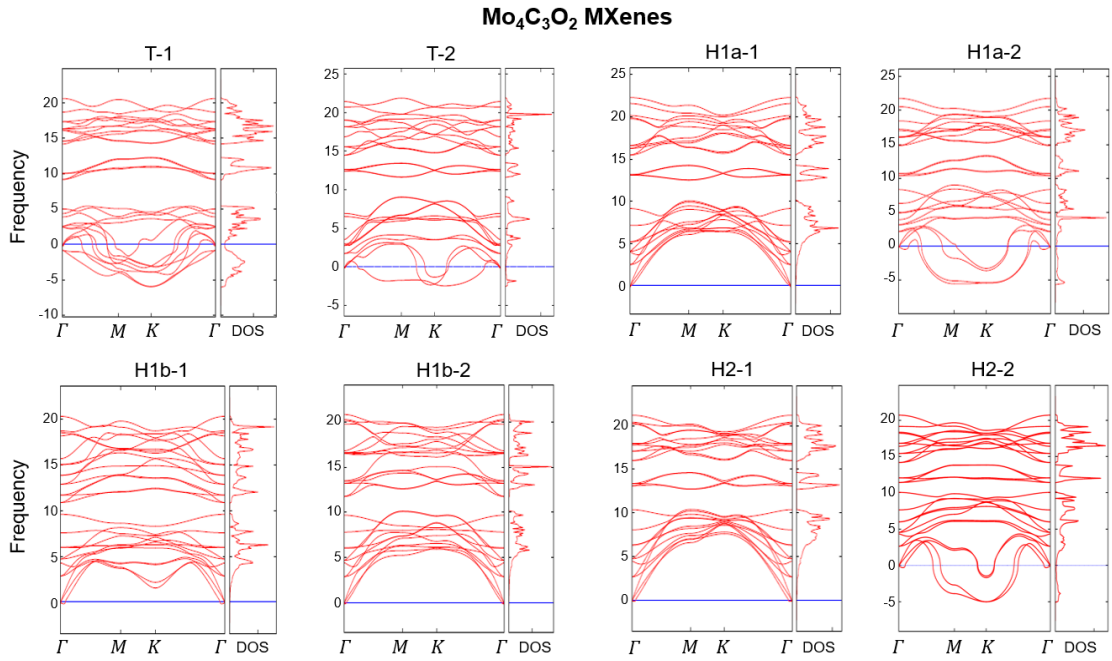

Fig. S19. Phonon spectra and vibrational density of states (DOSs) of eight types of Mo₄C₃O₂ MXenes terminated by oxygen atoms.

**Mo₄N₃O₂ MXenes**

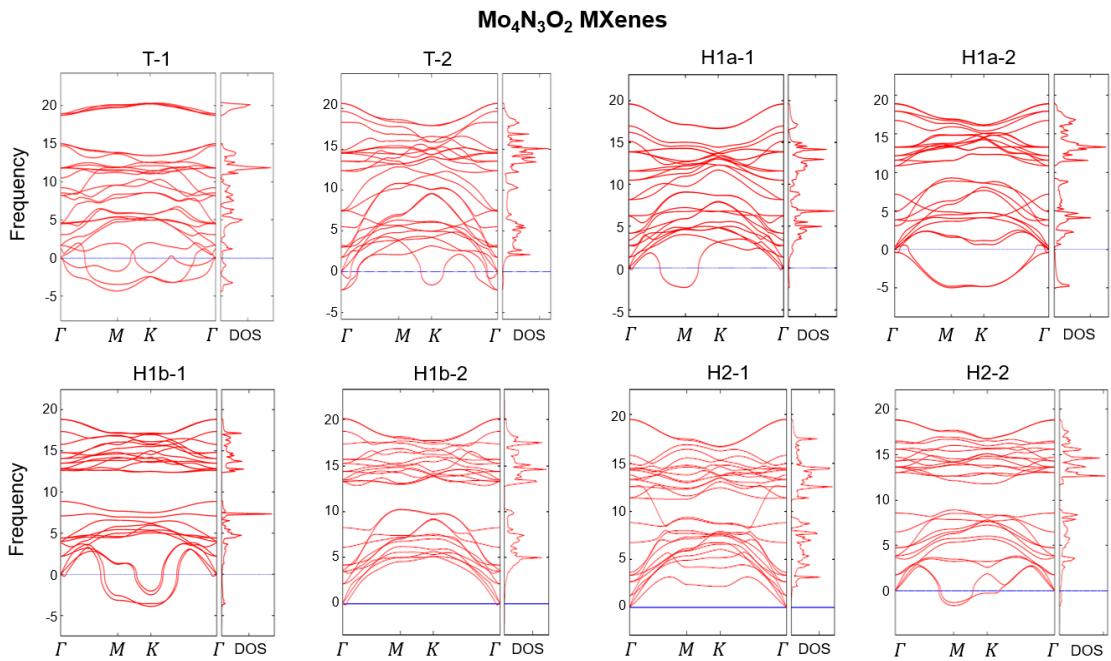

Fig. S20. Phonon spectra and vibrational density of states (vDOS) of eight types of Mo₄N₃O₂ MXenes terminated by oxygen atoms.

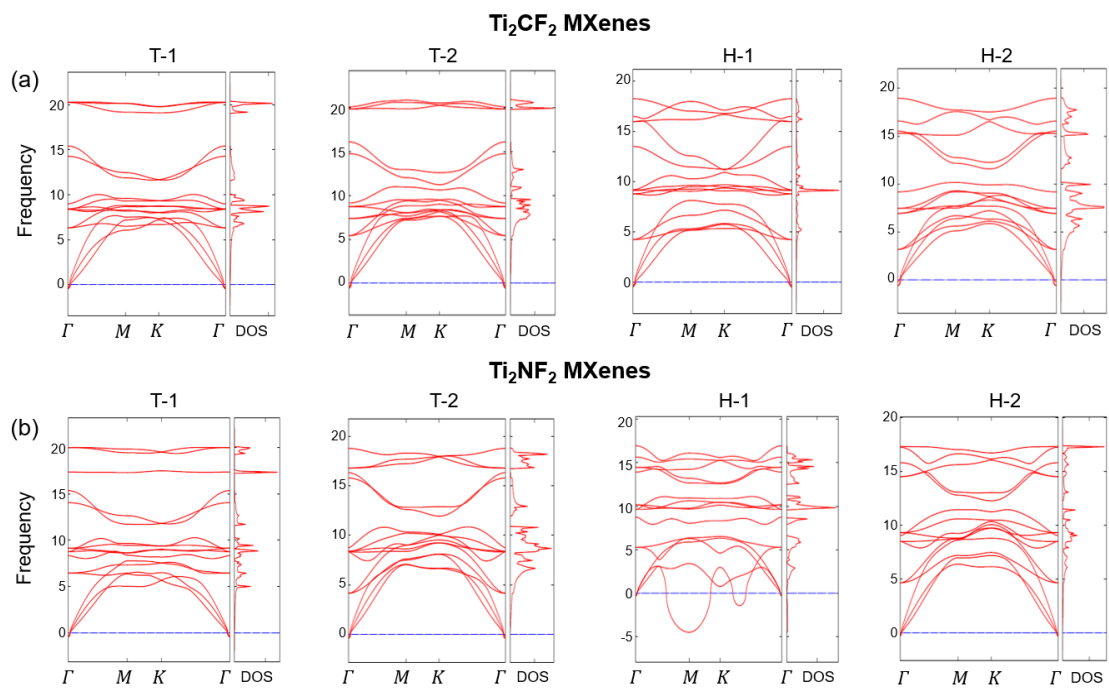

Fig. S21. Phonon spectra and vibrational density of states (vDOS) of (a) four $Ti_2CF_2$ MXenes and (b) four $Ti_2NF_2$ MXenes, both terminated by fluorine atoms.

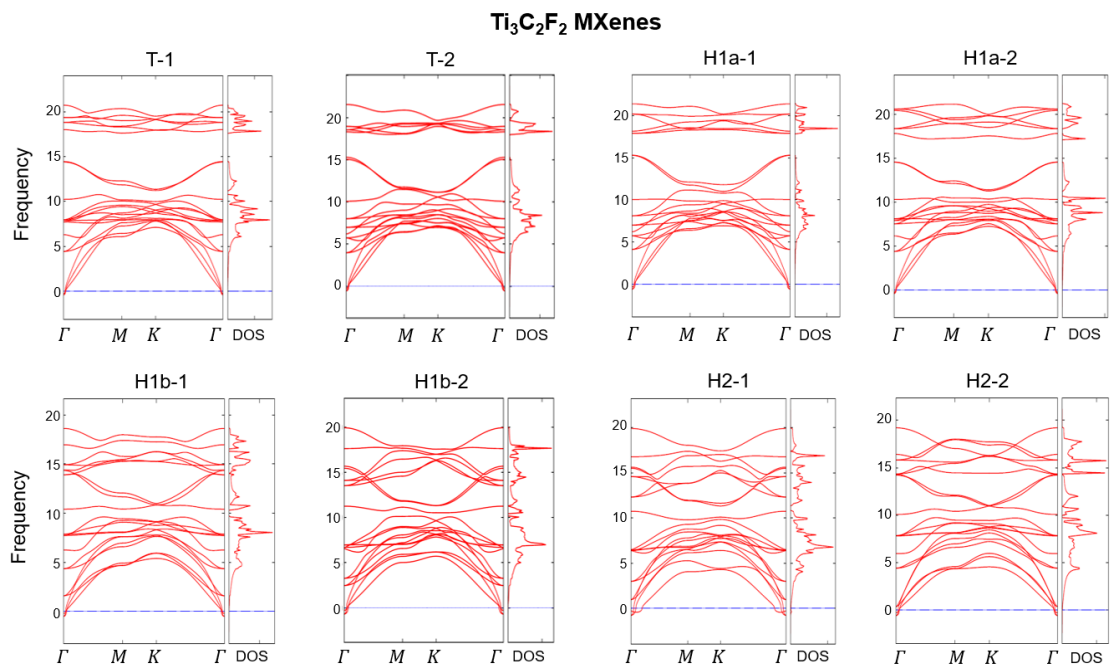

Fig. S22. Phonon spectra and vibrational density of states (vDOS) of eight types of Ti$_3$C$_2$F$_2$ MXenes terminated by fluorine atoms.

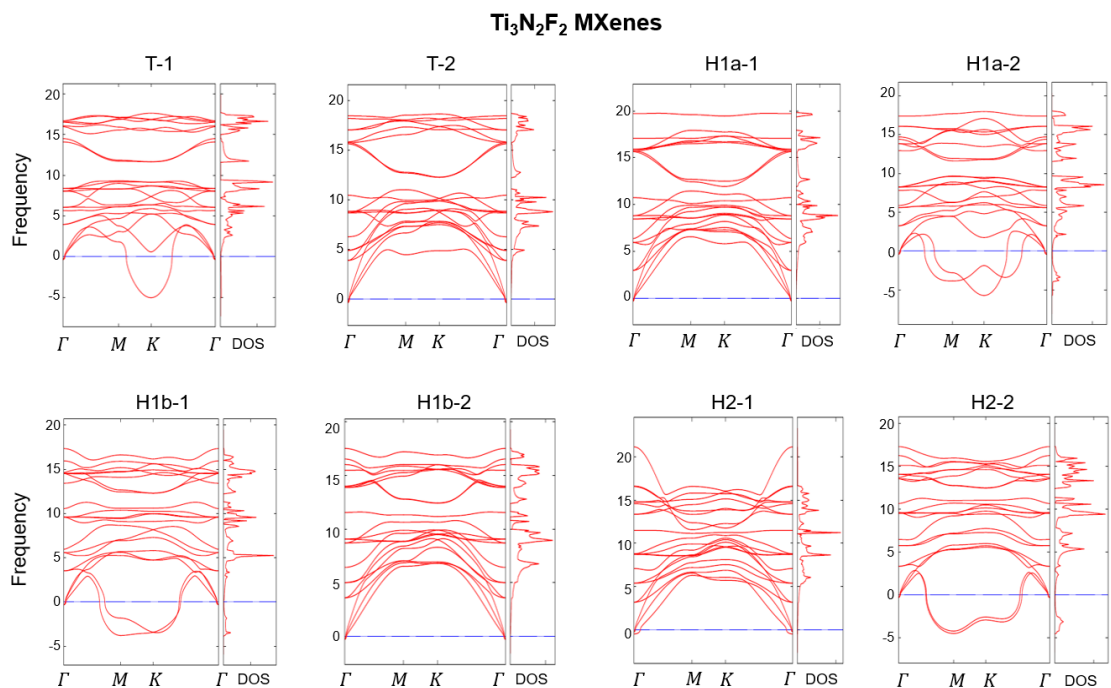

Fig. S23. Phonon spectra and vibrational density of states (vDOS) of eight types of Ti$_3$N$_2$F$_2$ MXenes terminated by fluorine atoms.

**Ti₄C₃F₂ MXenes**

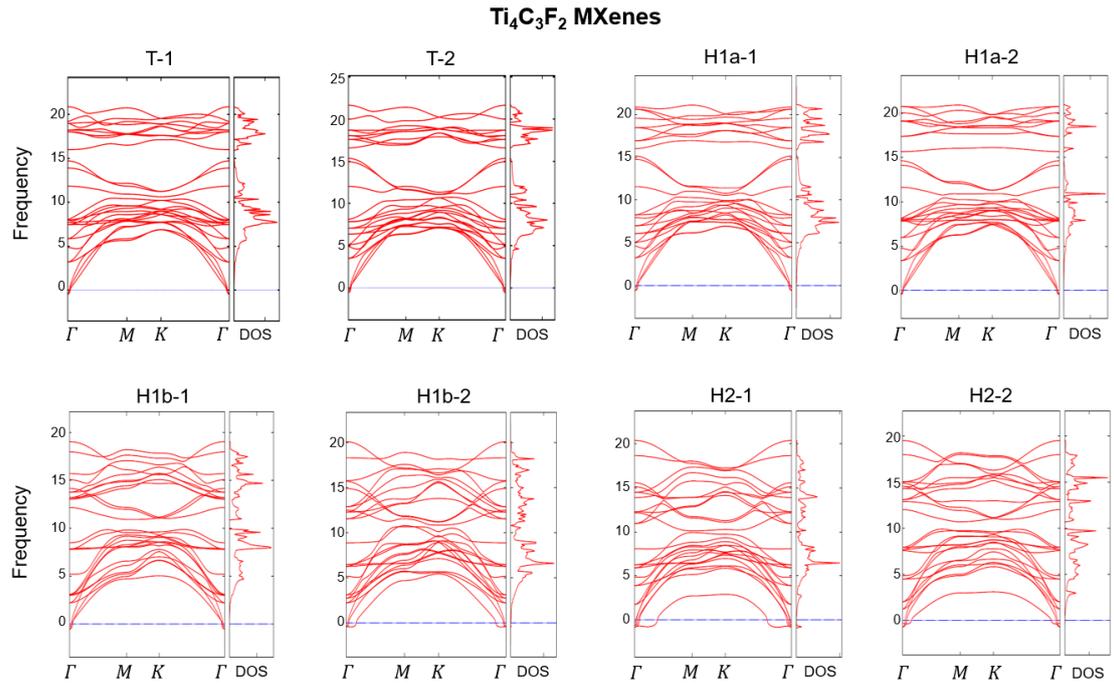

Fig. S24. Phonon spectra and vibrational density of states (vDOS) of eight types of Ti₄C₃F₂ MXenes terminated by fluorine atoms.

**Ti₄N₃F₂ MXenes**

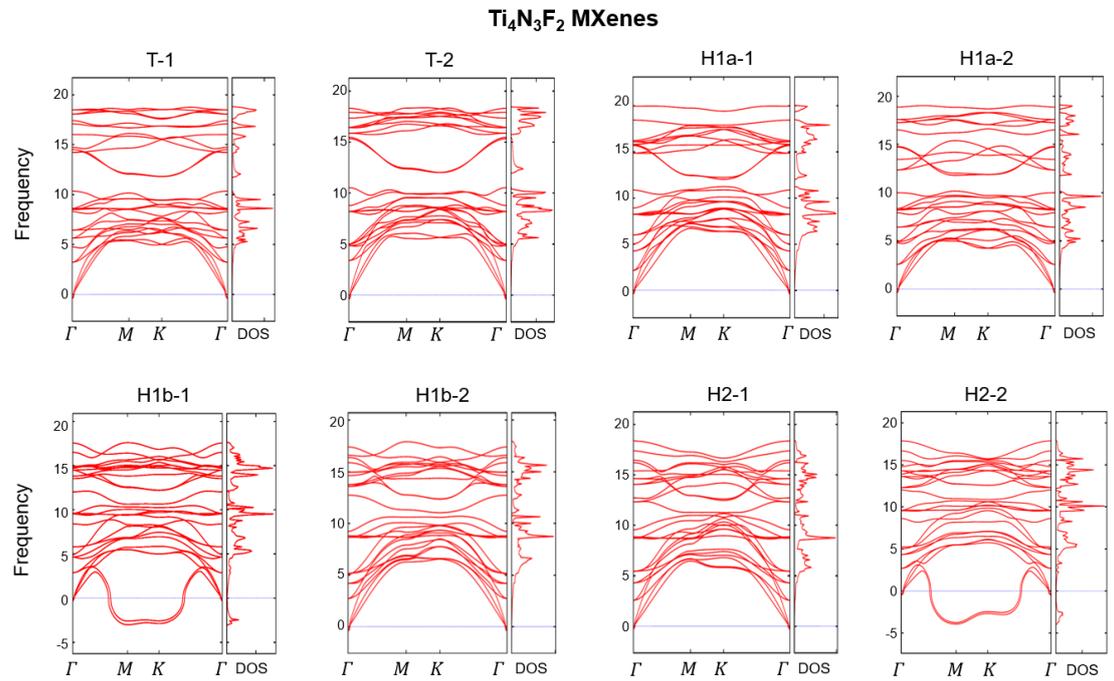

Fig. S25. Phonon spectra and vibrational density of states (vDOS) of eight types of Ti₄N₃F₂ MXenes terminated by fluorine atoms.

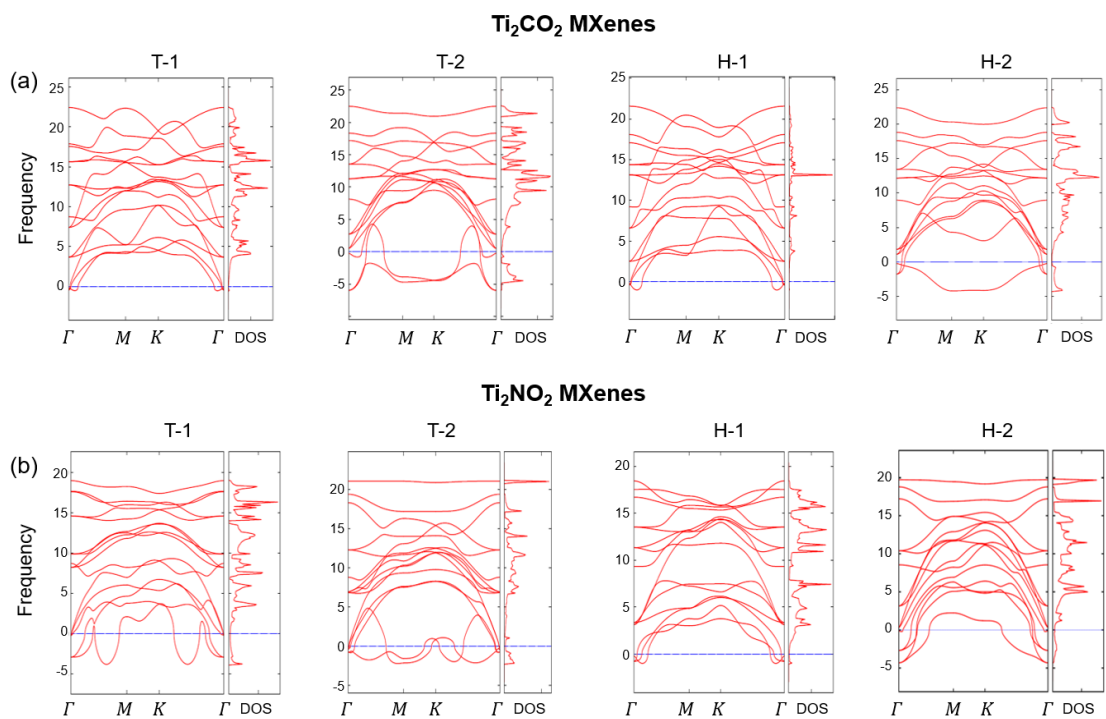

Fig. S26. Phonon spectra and vibrational density of states (vDOS) of (a) four Ti$_2$CO$_2$ MXenes and (b) four Ti$_2$NO$_2$ MXenes, both terminated by oxygen atoms.

**Ti₃C₂O₂ MXenes**

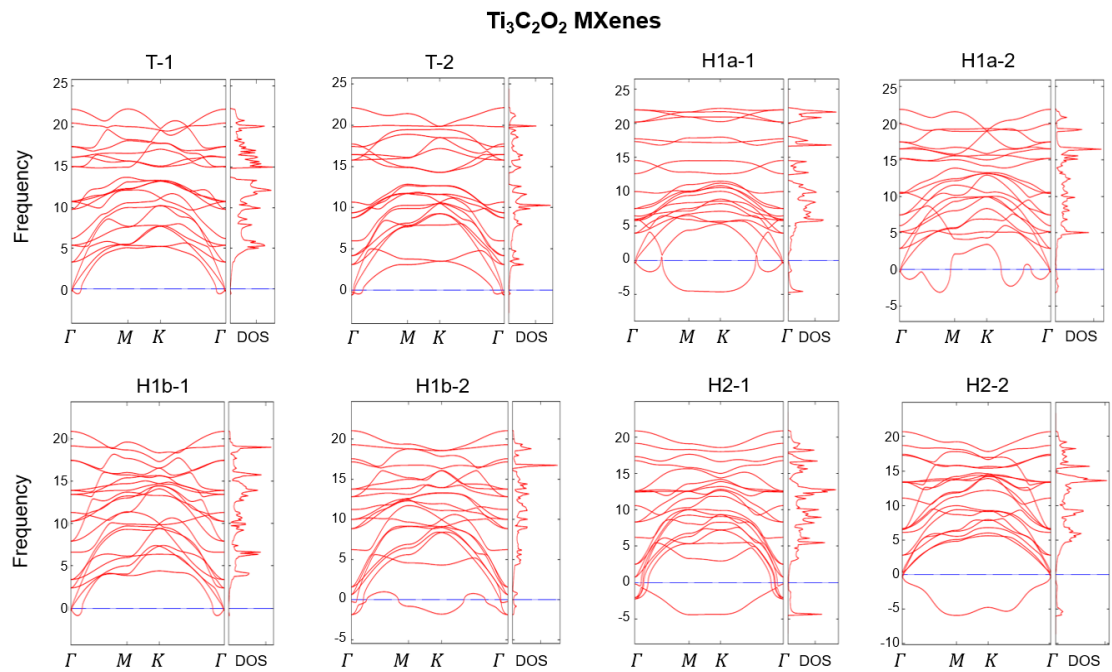

Fig. S27. Phonon spectra and vibrational density of states (vDOS) of eight types of Ti₃C₂O₂ MXenes terminated by oxygen atoms.

**Ti₃N₂O₂ MXenes**

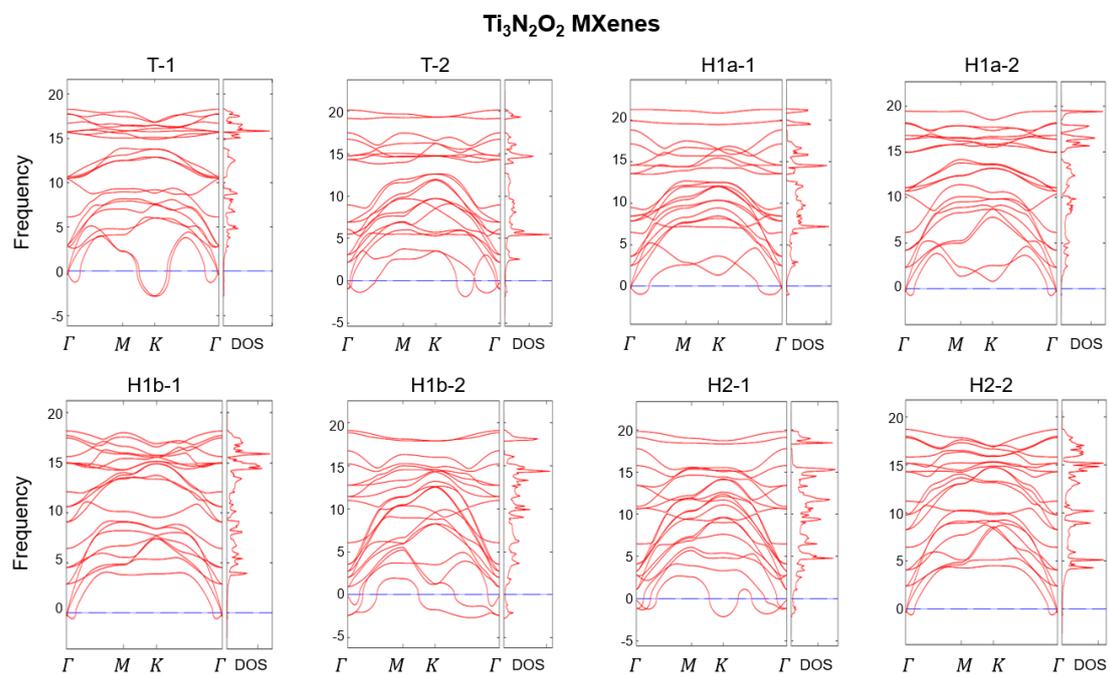

Fig. S28. Phonon spectra and vibrational density of states (vDOS) of eight types of Ti₃N₂O₂ MXenes terminated by oxygen atoms.

**Ti₄C₃O₂ MXenes**

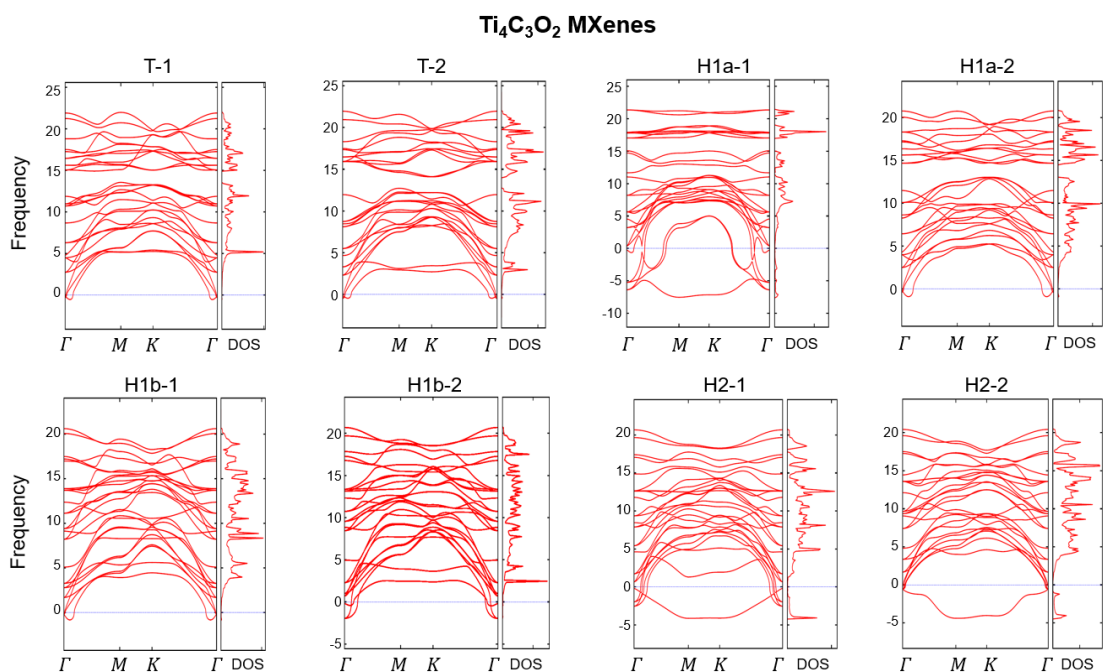

Fig. S29. Phonon spectra and vibrational density of states (vDOS) of eight types of Ti₄C₃O₂ MXenes terminated by oxygen atoms.

**Ti₄N₃O₂ MXenes**

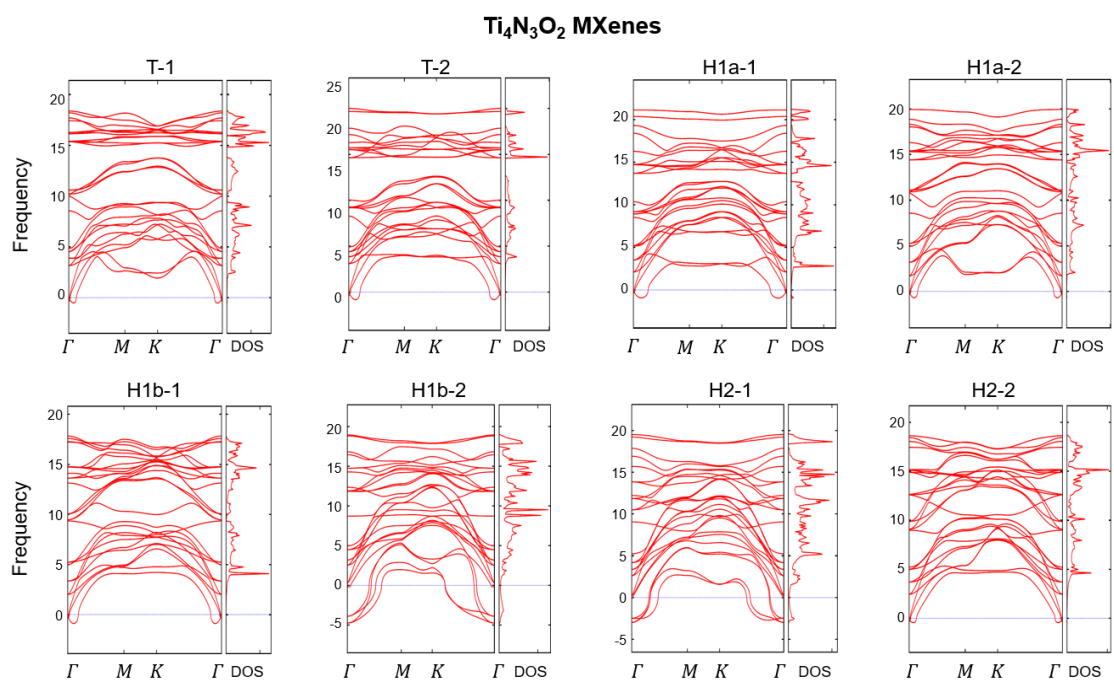

Fig. S30. Phonon spectra and vibrational density of states (vDOS) of eight types of Ti₄N₃O₂ MXenes terminated by oxygen atoms.